\documentclass[aps,prl,reprint,showpacs,floatfix]{revtex4-2}
\usepackage{amsmath,amsfonts,amssymb,graphics,graphicx,epsfig,color,times,indentfirst,layout,hyperref,subfigure,multirow,bm}
\usepackage{hyperref, soul}
\usepackage{ulem}
\hypersetup{linktocpage,,colorlinks=true}
\usepackage{threeparttable}
\usepackage{dcolumn}
\usepackage{bm}
\usepackage{color}

\date{\today}

\begin{document}
	
\title{$Z_2$ topological signatures of the optical bound on maximal Berry curvature: Application to two-dimensional time-reversal symmetric insulators}
\author{Pok Man Chiu}\email{pmchiu2022@gmail.com}
\affiliation{Graduate Institute of Applied Physics, National Chengchi University, Taipei City 11605, Taiwan}
\affiliation{Department of Physics, National Tsing Hua University, Hsinchu 30013, Taiwan}

\begin{abstract}
Unlike broken time-reversal symmetric (TRS) systems with a well-defined Chern number, directly measuring the bulk $Z_{2}$ invariant and Berry curvature (if nonzero) in topological insulators and their higher-order topological families remains an unsolved problem. Here, based on the refined trace-determinant inequality (TDI) involving the trace and determinant of the quantum metric and maximal Berry curvature (MBC), we propose an optical bound on the MBC for two-dimensional TRS insulators. As a result, using experimental data from a series of measurements—where the band‑inversion parameter is tuned and the optical conductivity is measured over a certain energy range—one can identify the $Z_2$ topological signature and construct the topological phase diagram by integrating the optical bound over frequency. This is supported by the momentum integration of the refined TDI, $f$-sum rule, and its topological extension, which provide a topological lower bound. To clearly identify the $Z_2$ topological signature, the faster decay of the optical weight in the topologically trivial region is crucial; this faster decay can be controlled by the optical gap and the inverse mass tensor. Crucially, the MBC we introduced enables us to prove the existence of tight topological lower bounds for the optical weight, quantum weight, and the double quantum volume. Based on the tight topological lower bounds, their physical meaning can be interpreted as an upper bound on the number of boundary states. We illustrate our approach using three representative topological models: the Kane-Mele model, mirror-protected insulator, and quadrupole insulator. Our results demonstrate that the MBC can reveal symmetry-protected topology and plays a role analogous to that of the original Berry curvature. Our framework opens an avenue for extracting the $Z_2$ topological signature of TRS insulators from optical conductivity data.
\end{abstract}

\maketitle




\textit{Introduction}.---In the presence or absence of time-reversal symmetry, topological insulating phases can be categorized into two distinct groups: TRS phases and magnetic phases. The former group includes topological insulators (TIs) \cite{Hasan10,Qi11,Ando13}, topological crystalline insulators (TCIs) \cite{Ando15,Chiu16}, and higher-order topological insulators (HOTIs) \cite{Yang24}. When time-reversal symmetry is broken, the resulting phases include Chern insulators \cite{Haldane88,Chang23} and fractional Chern insulators (FCIs) \cite{Parameswaran13,Bergholtz13,Neupert15,Bergholtz24,Park23,Cai23,Zeng23,Xu23,Ji24,Anderson24,Redekop24}. These time-reversal broken phases share a common characteristic: they possess a nonzero (fractional) Chern number. However, even when time-reversal symmetry is broken, there are topological phases that exhibit no bulk Chern number, such as axion insulators \cite{Nenno20,Sekine21} and antiferromagnetic topological insulators \cite{Tokura19,Wang21,Bernevig22}, whose topology is described by the $\theta$-term \cite{Qi08} or $Z_{2}$ invariant \cite{Mong10}. Recently, several studies have focused on topological systems with a nonzero Chern number (or Euler number)  \cite{Tran17,Ozawa18,Repellin19,Pozo19,Ozawa19,Jankowski23,Lysne23,Kruchkov24,Ryu23b,Onishi24,Ghosh24,Komissarov24,Verma24,Verma24b,Qiu24,Bac25}. All these works have proposed using optical circular dichroism measurements to probe the Chern number. However, in TRS systems with certain crystalline symmetries \cite{Xiao10,Fang12,Ortix21b}, there are no bulk Chern numbers or Berry curvature. How does topology emerge with the vanishing of Berry curvature, and how can it be probed through optical measurements? This remains elusive.



In this paper, we propose a measurable optical bound derived from the refined TDI, which is related to the trace and determinant of the quantum metric and the MBC \cite{maximal,Hardy30,Grafakos04} in TRS systems. Based on the "topological inequality" we established, the optical bound provides a useful tool for experimentally identifying the $Z_2$ topological signature and probing the upper limit of MBC as a function of frequency in these systems. Here the $Z_2$ topological signature can be revealed through a TRS-enforced topological lower bound. This bound determines the number of symmetry-protected boundary states and directly relates to topological invariants such as the $Z_2$ invariant and the mirror Chern number. Specifically, if the frequency integration of the optical bound decays sufficiently rapidly in the topologically trivial region and exceeds the topological lower bound, then a topological phase can be identified. In this work, we use optical weight, quantum weight, quantum volume, and the projection form of MBC (defined below) to reveal the $Z_2$ topological signatures in TRS systems. In contrast, the Kubo form of MBC (defined below) can be used in TI/TCI or HOTI systems under certain conditions—for example, in the absence or presence of certain types of broken crystalline symmetry. Note that the MBC is a gauge invariant quantity. It is important to note that Berry curvature vanishes in the presence of both time-reversal symmetry and inversion symmetry (or mirror and rotational symmetry) \cite{Fang12}. To reveal how the vanishing of the Berry curvature is related to topology, we define the Kubo form of MBC by inserting an absolute‑value operator into the band summation of the Kubo formula expression for the Berry curvature, i.e.,
\begin{align}
	\overline{\Omega}_{ab}(\bm{k})=\sum_{m\in occ}\sum_{n\in unocc}\overline{\Omega}^{mn}_{ab}(\bm{k}),
\end{align}
where $\overline{\Omega}^{mn}_{ab}(\bm{k})=|-2\Im(r^{a}_{mn}(\bm{k})r^{b}_{nm}(\bm{k}))|$ and $r^{a}_{mn}(\bm{k})=\langle u_{m}(\mathbf{k})|i\partial_{\bm{k}_a}|u_{n}(\mathbf{k})\rangle$ is the interband Berry connection, and $a=x,y$. In most cases, the Kubo form of MBC does not vanish and is sensitive to certain types of broken crystalline symmetry in the lattice, which reduce its value and thus make it smaller than the topological lower bound. However, this definition is more directly related to optical measurements. To eliminate the influence of broken crystalline symmetry, we can also define the projection form of MBC using the projection operator as,
\begin{align}
	\overline{\overline{\Omega}}_{ab}(\mathbf{k})=\mathrm{tr}(|iP(\mathbf{k}) [\partial_a P(\mathbf{k}),\partial_b P(\mathbf{k})]|),
\end{align}
where $P(\mathbf{k})$ is the projection operator of occupied bands. Here we define $|X|=\sqrt{X^{\dagger}X}$, and therefore $\overline{\overline{\Omega}}_{ab}(\mathbf{k})$ can be viewed as the Schatten 1-norm of the operator $iP(\mathbf{k}) [\partial_a P(\mathbf{k}),\partial_b P(\mathbf{k})]$ at each momentum point. The advantage of this definition is that it preserves topological information and always remains larger than the tight topological lower bound. Crucially, a tight topological lower bound for TRS systems can be obtained from the maximal property of the MBC.

\textit{Kubo-Greenwood formula}.---The linear optical conductivity can be expressed by the Kubo-Greenwood formula \cite{Kubo57,Greenwood58,Huhtinen23}:
\begin{align}
	\sigma_{ab}(\omega)=\frac{e^{2}}{i\hbar}\sum_{m\ne n}\int\frac{d^2\bf{k}}{(2\pi)^2}\frac{f_{mn}(\mathbf{k})}{E_{mn}(\mathbf{k})}  
    \frac{\mathcal{M}^{a}_{mn}(\mathbf{k})\mathcal{M}^{b}_{nm}(\mathbf{k})}{\hbar\omega+E_{mn}(\mathbf{k})+i0^{+}},
\end{align}
where the velocity matrix element $\mathcal{M}^{a}_{mn}(\mathbf{k})=\langle u_{m}(\mathbf{k})|\hbar\hat{\upsilon}_{a}|u_{n}(\mathbf{k})\rangle$.  $f_{n}(\mathbf{k})=1/[1+e^{(E_{n}(\mathbf{k})-\mu)/k_{B}T}]$ and $\hat{\upsilon}_{a}=\frac{1}{\hbar}\frac{\partial \hat H}{\partial k_{a}}$ are the Fermi-Dirac distribution function and velocity operator, respectively. We denote $f_{mn}(\mathbf{k})=f_{m}(\mathbf{k})-f_{n}(\mathbf{k})$ and $E_{mn}(\mathbf{k})=E_{m}(\mathbf{k})-E_{n}(\mathbf{k})$, where $H(\mathbf{k})|u_{n}(\mathbf{k})\rangle=E_{n}(\mathbf{k})|u_{n}(\mathbf{k})\rangle$. Here $\mu$ is the chemical potential. In the presence of interactions, one may directly replace $H(\mathbf{k})$, $E_{n}(\mathbf{k})$ and $u_{n}(\mathbf{k})$
with their many-body counterparts under twisted boundary conditions \cite{Niu85,Souza00,Resta05,Onishi24} and then obtain the corresponding expressions for the optical conductivity and the quantum geometric tensor.

The real part of longitudinal optical conductivity corresponds to absorption, which is the primary focus of this work. The interband contribution to the real part of optical conductivity in the clean limit can be expressed as follows:
\begin{align}
	\Re\sigma_{aa}(\omega)&=-\frac{\pi e^{2}}{\hbar}\sum_{m\ne n}\int\frac{d^2\bf{k}}{(2\pi)^2}\frac{f_{mn}(\mathbf{k})}{E_{mn}(\mathbf{k})} 
	|\mathcal{M}^{a}_{mn}(\mathbf{k})|^2  \nonumber \\
	&\times\delta(\hbar\omega+E_{mn}(\mathbf{k})).
\end{align}
Since we consider the TRS systems, there is no optical Hall conductivity. To connect this with quantum geometry, we assume the system is at zero temperature and utilize the relationship between the velocity matrix element and the interband Berry connection, $\langle u_{m}(\mathbf{k})|\hbar\hat{\upsilon}_{a}|u_n(\bm{k})\rangle=i E_{mn}r^{a}_{mn}(\bm{k})$ where $m\ne n$, to express the real part of optical conductivity as
\begin{align}
	\Re\sigma_{aa}(\omega) 
    =\pi e^{2}\omega\sum_{m\ne n}\int\frac{d^2\bf{k}}{(2\pi)^2}
	r^{a}_{mn}r^{a}_{nm}\delta(\hbar\omega+E_{mn,\bm{k}}).
\end{align}
Furthermore, one can introduce the quantum geometric tensor \cite{Ahn20,Ahn22,Resta11}, $Q^{mn}_{ab}(\bm{k})=r^{a}_{mn}(\bm{k})r^{b}_{nm}(\bm{k})=g^{mn}_{ab}(\bm{k})-\frac{i}{2}\Omega^{mn}_{ab}(\bm{k})$,
where $g^{mn}_{ab}(\bm{k})$ and $\Omega^{mn}_{ab}(\bm{k})$ are the component of quantum metric and Berry curvature, respectively. Note that in linearly polarized optical responses, the term for Berry curvature is absent. Recall that there is no Berry curvature in TRS systems that possess inversion or other symmetries \cite{Ortix21b}. Finally, we can rewrite the absorption part of longitudinal conductivity as
\begin{align}
	\Re\sigma_{aa}(\omega)=\pi e^{2}\omega\sum_{m\ne n}\int\frac{d^2\bf{k}}{(2\pi)^2}g^{mn}_{aa}(\bm{k})\delta(\hbar\omega+E_{mn}(\mathbf{k})).
\end{align}
In this work, we mainly focus on a quantity, which we call optical bound, i.e.,
\begin{align}
	\mathcal{K}_{OP}(\omega)=\frac{2[\Re\sigma_{xx}(\omega)+\Re\sigma_{yy}(\omega)]}{\hbar\omega},
\end{align}
whose frequency integration is given by $\mathcal{K}^{\infty}_{OP}=\frac{\hbar}{2\pi}\int^{\infty}_{0}d\omega\mathcal{K}_{OP}(\omega)$, which contains the topological signatures provided by the Souza-Wilkens-Martin (SWM) sum rule \cite{Souza00}, the refined TDI and its topological extension defined below. To reveal the behavior of the MBC, we introduce the corresponding imaginary part of the generalized (circularly polarized) optical Hall conductivity \cite{Ebert96,Oppeneer98,Gradhand13,Onishi24,Bac25}, i.e.,
\begin{align}
	\Im\overline{\sigma}_{xy}(\omega)=\frac{\pi e^{2}}{\hbar}\sum_{m\ne n}\int\frac{d^2\bf{k}}{(2\pi)^2}\overline{\Omega}^{mn}_{xy}(\bm{k})\delta(\hbar\omega+E_{mn}(\mathbf{k})).
\end{align}
The magnitude of $\Im\overline{\sigma}_{xy}(\omega)$ is always lower than the optical bound, which we will study below.

\begin{figure*}[]
	\includegraphics[width=0.99\textwidth]{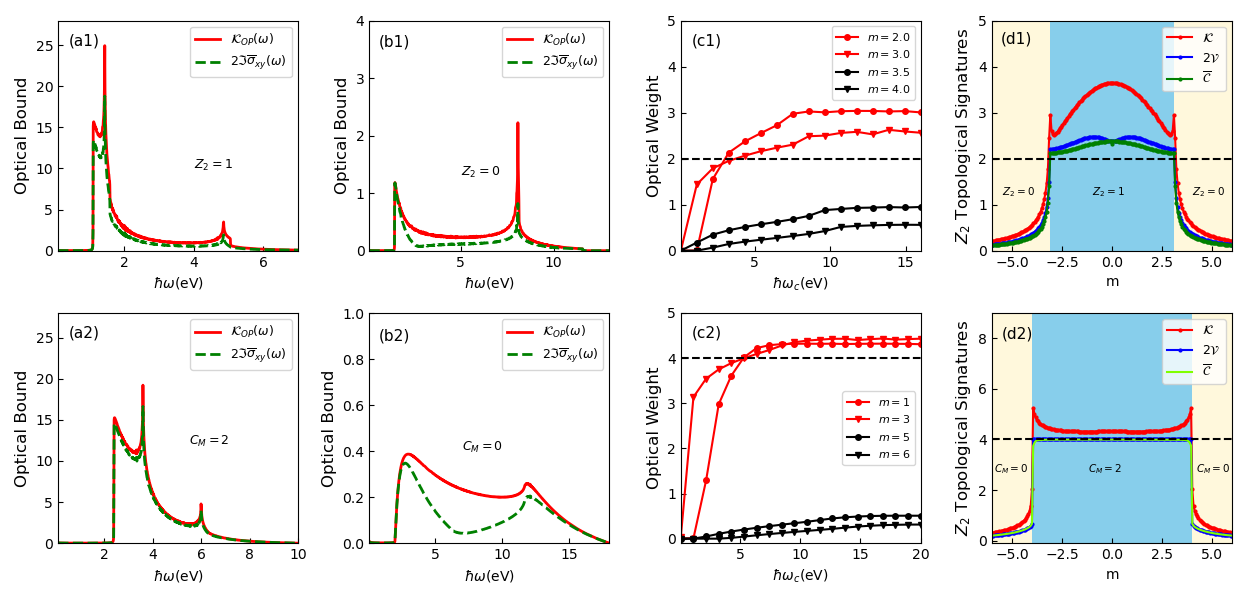}
	\caption{Optical bound, optical weight, and $Z_2$ topological signatures of the Kane-Mele model and TCI model of mirror-protected insulator. (a1, a2, b1, b2) The optical bound as a function of frequency with $\lambda_{\upsilon}=2t,4t$ in (a1,b1) and $\upsilon_{m}=1,5$ in (a2,b2). The red solid (green dashed) line represents the optical bound (generalized optical Hall conductivity) in the nontrivial (a1, a2) and trivial phases (b1, b2). (c1, c2) The optical weight $\mathcal{K}^{\omega_c}_{OP}$ as a function of frequency cutoff for four parameters. The red (black) lines correspond to the nontrivial (trivial) phase. (d1, d2) The $Z_2$ topological signatures $\mathcal{K}$, $2\mathcal{V}$ and $\overline{C}$ as a function of a range of parameters. The back dashed line represents the number of boundary states. We calculate $\Re\sigma_{aa}$ and $\Im\overline{\sigma}_{ab}$ in units of $e^2/h$, and $\overline{C}$ using the Kubo form of the MBC. Here, we set $(t,\lambda_{SO},\lambda_{R},\lambda_{\upsilon})=(1,0.6t,0,m)$ and $(t_1,t_2,t_{PH},\upsilon_s,\upsilon_{m})=(2,1.5,0.1,1.3,m)$ for the Kane-Mele model and TCI \cite{Wieder20}, respectively. 
	}
\end{figure*}

\textit{Refined trace-determinant inequality}.---The refined TDI serves as a foundational element for establishing the concept of optical bounds. It links geometric and topological information with experimental observables. To begin, we will first review the trace condition \cite{Roy14} in two dimensions: $\mathrm{tr} g({\bf k}) \ge |\Omega({\bf k})|$, where $\mathrm{tr} g({\bf k})=g_{xx}({\bf k})+g_{yy}({\bf k})$ is the trace of non-Abelian quantum metric \cite{Rezakhani10,Ma10,Ahn20}, and $\Omega({\bf k})$ is the non-Abelian Berry curvature. The TDI was first developed for flat Chern bands in lattice systems as a condition to satisfy the Girvin-MacDonald-Platzman algebra \cite{GMP86} of projected density operators to all orders, with the aim of realizing FCI phases \cite{Parameswaran13,Bergholtz13,Neupert15,Bergholtz24}. When the trace condition is saturated, the FCI phases can be achieved in lattice systems. Since we consider the TRS systems, both the Berry curvature and the Chern number are zero. The determinant condition \cite{Roy14,det}, i.e., $\sqrt{\det(g({\bf k}))} \ge |\Omega({\bf k})|/2$, does not yield useful information. Instead, to combine the two conditions, we propose the following refined TDI:
\begin{align}
	\label{eq:e1}
	\mathrm{tr} g({\bf k}) \ge 2\sqrt{\det(g({\bf k}))} \ge \overline{\Omega}_{ab}({\bf k}).
\end{align}
where $\overline{\Omega}_{ab}({\bf k})$ is the MBC characterizing the nontrivial topology of a TRS system. Note that $\overline{\Omega}_{ab}({\bf k})\ge|\Omega_{ab}({\bf k})|$. Crucially, based on the semi-positive definite property of the quantum metric, this refined TDI holds for arbitrary quantum states in two dimensions. When we integrate the refined TDI both sides, the inequality becomes:
\begin{align}
	\int\frac{d^2\mathbf{k}}{2\pi}\mathrm{tr} g({\bf k}) \ge \int\frac{d^2\mathbf{k}}{2\pi}2\sqrt{\det(g({\bf k}))} \ge \int\frac{d^2\mathbf{k}}{2\pi}\overline{\Omega}_{ab}({\bf k}).
\end{align}
In the following discussion, we denote the above integrated TDI as $\mathcal{K}\ge2\mathcal{V}\ge\overline{C}$. Importantly, $\mathcal{K}$ called quantum weight can be observed through optical responses, whereas $\mathcal{V}$ and $\overline{C}$ represents the quantum volume \cite{Mera21,Ozawa21,Mera21b} and "pseudo-Chern number" of the occupied bands, respectively. Similarly, we define the "pseudo-Chern number" of the projection form of MBC as $\overline{\overline{C}}$. Note that after multiplying by the normalized constant $\frac{1}{2\pi}$, the quantum volume has a transparent physical interpretation. Recall that in TRS systems, such as TIs, TCIs, and HOTIs, there is no bulk Chern number or quadrupole moment; however, a nonzero $\overline{C}$ can still exist and will be slightly larger than or equal to the number of boundary states, provided that no term breaks certain crystalline symmetries or that the system undergoes specific topological deformations. To reveal topological information faithfully even in the presence of broken crystalline symmetry, we prove the existence of a tight topological lower bound using the projection method \cite{Kivelson82,Prodan09,Prodan10,Sheng11,Li13,Lin24}, which will be discussed below.


As our main result for TI/TCI systems, the decay of quantum geometric quantities in the topologically trivial region and their variation during topological deformation are constrained by the following  "topological inequality" \cite{Onishi24,supp}:
\begin{align}
	\frac{\pi e^2}{2}\frac{\mathcal{M}}{E_g}\ge\mathcal{K}\ge2\mathcal{V}\ge\overline{\overline{C}}\ge\sum_{s=1}^{n_t}|\mathcal{N}_{s}|
\end{align}
where $\mathcal{M}=\sum_{m,n,a}\int \frac{d\mathbf{k}}{(2\pi)^2}[M^{-1}_{mn}(\mathbf{k})]_{aa}\langle c^{\dagger}_{\mathbf{k}m}c_{\mathbf{k}n}\rangle$ and  $[M^{-1}_{mn}(\mathbf{k})]_{aa}=\sum_{\alpha\beta}U^{\dagger}_{m,\alpha}(\mathbf{k})\frac{\partial^2H_{\alpha\beta}(\mathbf{k})}{\partial(\hbar k_a)^2}U_{\beta,n}(\mathbf{k})$ is the inverse mass tensor \cite{Kubo57,Hazra19}. The topological number $\mathcal{N}_{s}$ will be defined below. Here $E_g$ is the optical gap \cite{Onishi24}. In other words, the optical gap and the degree of flatness \cite{Chiu25}, which relates to the inverse mass tensor \cite{Onishi24, supp}, are the two key factors controlling the variation of the quantum weight and the quantum volume. Crystalline symmetry also significantly affects the decay of quantum weight in the topologically trivial region (see below and Appendix A).

In TRS systems, the total Chern number of the occupied bands is always zero. However, one can project the occupied bands onto different topological sections \cite{Prodan09,Prodan10,Sheng11,Li13,Lin24}. As a result, each topological sector has a well-defined Chern number. Here, we use the spectral theorem \cite{Goldbring21} or the spectral representation \cite{Kreyszig78} to reveal the relationship between $P(\mathbf{k})$ before and after projecting onto different topological sections, i.e., $P(\mathbf{k})=\sum^{occ}_{n=1}|u_{n}(\mathbf{k}) \rangle \langle u_{n}(\mathbf{k})|=\sum^{N_+}_{i=1}|\psi^+_i(\mathbf{k}) \rangle \langle \psi^+_i(\mathbf{k})|+\sum^{N_-}_{i=1}|\psi^-_i(\mathbf{k}) \rangle \langle \psi^-_i(\mathbf{k})|$. This can be realized as the projected bands $|\psi^\pm_i(\mathbf{k}) \rangle$ , which belong to two topological sections, being subject to two constraints \cite{Kivelson82,Lin24}, i.e., $P(\mathbf{k})|\psi^\pm_i(\mathbf{k})\rangle=|\psi^\pm_i(\mathbf{k})\rangle$ and $P(\mathbf{k})\hat{s}P(\mathbf{k})|\psi^\pm_i(\mathbf{k})\rangle=\lambda^\pm_i|\psi^\pm_i(\mathbf{k})\rangle$, where $\hat{s}$ is a suitable operator that can isolate the topology in different sectors. In other words, $P(\mathbf{k})$ has two different sets of eigenstates $\{|\psi^\pm_i(\mathbf{k}) \rangle\}$
and $\{|u_{n}(\mathbf{k}) \rangle\}$, which represent the same system. In fact, $|\psi^\pm_i(\mathbf{k}) \rangle$ is simply a linear combination of $|u_{n}(\mathbf{k}) \rangle$. Consequently, the integrated TDI Eq. (9) is extended as follows:
\begin{align}
	\mathcal{K}\ge2\mathcal{V}\ge\overline{\overline{C}}\ge\sum_{s=1}^{n_t}|\mathcal{N}_{s}|,
\end{align}
where $n_t$ and $\mathcal{N}_{s}$ represent the number of topological sections and the topological number in different sections (see SM \cite{supp} for more details), respectively. In general, except for the (pseudo)spin Chern number, we argue that $\mathcal{N}_{s}$ can be the quadrupole moment, Wannier-band polarization, or other topological invariants. In this work, we provide a proof only for the case of the (pseudo)spin Chern number. Here two points are worth emphasizing: (i) the topological number $\mathcal{N}_s$ is nonzero only in projected bands, whereas the quantum weight and the quantum volume are the same for both the original and projected bands; (ii) Eqs. (10), (11) and (12) should be viewed in the same basis, namely original or projected bands. We performed all computations using the original bands. For TI/TCI systems, one can utilize the projected spin operator, $P({\bf k})\hat{s}P({\bf k})$ \cite{Prodan09,Prodan10,Sheng11,Li13,Lin24}, to separate the occupied bands into spin-up and spin-down sections, each of which has a well-defined Chern number. Here $\hat{s}$ is the (pseudo)spin operator.  As a result, $\mathcal{N}_{s}$ becomes the (pseudo)spin Chern number $C_{\pm}$ \cite{Prodan09,Prodan10,Sheng11,Li13,Lin24} in the case of TIs/TCIs. Since the topology of HOTIs is characterized by the Wannier-band polarization and quadrupole moment \cite{Benalcazar17,Benalcazar17b}, the Kubo form of MBC manifests a topological lower bound only through specific topological deformations, but it generally captures the pattern of the hidden Berry curvature in HOTI systems. Importantly, the inequality $\mathcal{K}\ge2\mathcal{V}\ge\overline{C}$ always holds for HOTIs. It is important to note that the quantum metric is a robust geometric quantity that is always non-vanishing. In contrast, topological invariants such as the Chern number and the $Z_2$ invariant vanish in TIs and HOTIs, respectively, and cannot be measured directly. Fortunately, by utilizing data from optical response measurements—specifically, the optical bound—one can obtain topological signals, which we will explore further below.


\begin{figure*}[]
	\includegraphics[width=0.99\textwidth]{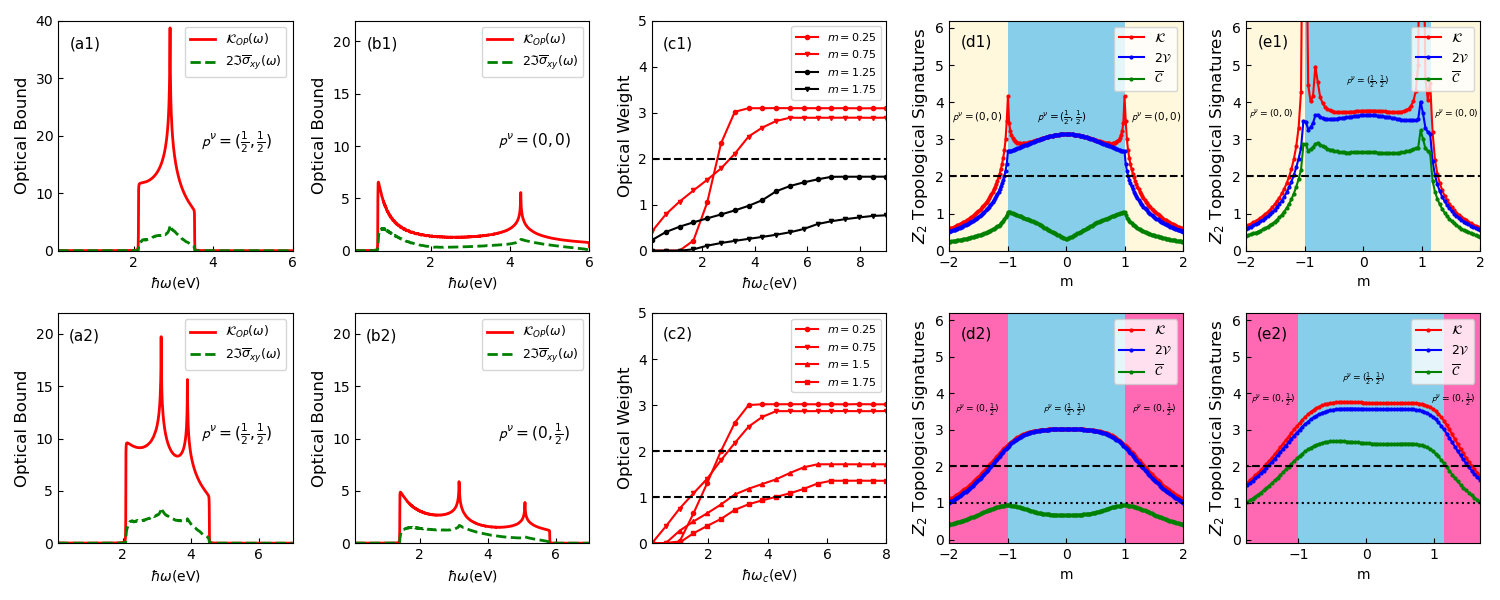}
	\caption{Optical bounds, optical weight, and $Z_2$ topological signatures of HOTI model of quadrupole  insulator. (a1, a2, b1, b2) The optical bound as a function of frequency. The red solid (green dashed) line represents the optical bound (generalized optical Hall conductivity). In (a1, b1) and (a2, b2), we set $(\gamma_x, \gamma_y) = (0.25, 0.25)$, $(1.25, 1.25)$ and $(\gamma_x, \gamma_y) = (0.25, 0.5)$, $(1.5, 0.5)$, respectively. (c1, c2) The optical weight $\mathcal{K}^{\omega_c}_{OP}$ as a function of frequency cutoff for four parameters. The red (black) line corresponds to the nontrivial (trivial) phase. (d1, d2, e1, e2) The $Z_2$ topological signatures $\mathcal{K}$, $2\mathcal{V}$, and $\overline{C}$ as a function of a range of parameters without and with breaking mirror symmetry \cite{Li20}, respectively. In (e1, e2), we set the magnitude of the mirror-symmetry-breaking term to 0.62. Note that in the presence of the mirror-symmetry-breaking term, the topological phase diagram shows a slight shift as a function of $m$, and a slightly asymmetric behavior when changing the sign of $m$. The black dashed and dotted lines represent the number of boundary states in the phases $\mathbf{p}^{\nu} = \left(\frac{1}{2}, \frac{1}{2}\right)$ and $\mathbf{p}^{\nu} = \left(0, \frac{1}{2}\right)$, respectively. We calculate $\Re\sigma_{aa}$ and $\Im\overline{\sigma}_{ab}$ in units of $e^2/h$, and $\overline{C}$ using the Kubo form of the MBC. We take $(\gamma_x,\gamma_y) = (m,m)$ for panels (c1,d1) and $(\gamma_x,\gamma_y) = (m,0.5)$ for panels (c2,d2), respectively. Here, we set $(\lambda_x,\lambda_y)=(1,1)$.
	}
\end{figure*}

\textit{Optical bound}.---In the previous subsection, we discussed the refined TDI and the "topological inequality". Now, we will examine their relationship with optical responses and reveal their topological signatures. By utilizing the identities $r^{a}_{mn}=\hbar\upsilon^{a}_{mn}/(iE_{mn})$ and $|\upsilon^{x}_{mn}\pm i\upsilon^{y}_{mn}|^{2}=|\upsilon^{x}_{mn}|^{2}+|\upsilon^{y}_{mn}|^{2}\mp i(\upsilon^{x}_{mn}\upsilon^{y}_{nm}-\upsilon^{y}_{mn}\upsilon^{x}_{nm})$ which is always semi-positive \cite{Sekh22}, and by applying the TDI, we can derive the optical bound on the components of the MBC for any interband transition contour (see SM \cite{supp}), i.e.,
\begin{align}
	\frac{\Re\sigma^{xx}_{mn}(\mathbf{k},\omega)+\Re\sigma^{yy}_{mn}(\mathbf{k},\omega)}{\hbar\omega}\ge\Im\overline{\sigma}^{xy}_{mn}(\mathbf{k},\omega),
\end{align}
where we define $\Re\sigma^{aa}_{mn}(\mathbf{k},\omega)=c|\hbar\upsilon^{aa}_{mn}|^{2}\delta(\hbar\omega+E_{mn})/|E_{mn}|$, and $\Im\overline{\sigma}^{xy}_{mn}(\mathbf{k},\omega)=c|-2\Im(\hbar\upsilon^{x}_{mn}\hbar\upsilon^{y}_{nm}/E^{2}_{mn})|\delta(\hbar\omega+E_{mn})$. The constant, $c=\pi e^2/\hbar$. The corresponding real or imaginary part of $\sigma^{ab}_{mn}(\mathbf{k},\omega)$ which are not considered in this study, can also be defined.

After integrating the momentum variable and performing the band summation, we ultimately obtain the optical bound on MBC at any frequency,
\begin{align}
	\mathcal{K}_{OP}(\omega)\ge2\Im\overline{\sigma}_{xy}(\omega).
\end{align}
Here, we multiply both side of $2$ to maintain consistency with the definitions of quantum geometry and topology. Importantly, the left-hand term, which we refer to as the optical bound, can be obtained through measurements of optical responses. In contrast, the right-hand term represents the corresponding generalized optical Hall conductivity. To identify the range of frequencies that contribute to the $Z_2$ topological signature, we can calculate the optical weight as a function of the frequency cutoff $\omega_c$: $\mathcal{K}^{\omega_c}_{OP}=\frac{\hbar}{2\pi}\int^{\omega_c}_{0}d\omega\mathcal{K}_{OP}(\omega)$. By combining the two formulas $\mathcal{K}^{\infty}_{OP}=\frac{e^2}{h}\mathcal{K}$ (see SM \cite{supp}) and $\mathcal{K}\ge2\mathcal{V}\ge\overline{\overline{C}}\ge\sum_{s=1}^{n_t}|\mathcal{N}_{s}|$, one can extract the $Z_2$ topological signature from the optical weight $\mathcal{K}^{\infty}_{OP}$ if the quantum weight decays rapidly in the topologically trivial region. As a result, if the optical weight $\mathcal{K}^{\omega_c}_{OP}$ is larger than the number of boundary states, it carries $Z_2$ topological signature. For example, in the case of TIs/TCIs, the number of boundary states is $|C_+|+|C_-|$. If $\mathcal{K}^{\omega_c}_{OP}>|C_+|+|C_-|$, it gives a $Z_2$ topological signature. This always holds for a topologically nontrivial phase. However, near the topological transition point but on the topologically trivial side, the "geometric effect" of the dispersive bands can also make $\mathcal{K}^{\omega_c}_{OP}$ larger than the number of boundary states when the optical gap is small. Therefore, controlling the size of the optical gap can help to identify the topology when the system is close to the transition. Note that the (pseudo)spin Chern number in different sectors can be used to define the $Z_2$ invariant as $\Delta=|C_{\pm}|\mod 2$ for TIs, whereas $C_{\pm}$ can be odd or even for TCIs. This is consistent with the definition of the $Z_2$ invariant in Refs. \cite{Roy09a,Roy09b}.



\textit{Models and results}.---To reveal the topological and geometric information of the optical bound and TDI, we examine three typical TRS models: the Kane-Mele model, mirror-protected insulator, and quadrupole insulator. First, we consider the Kane-Mele model \cite{Kane05}, whose Hamiltonian is given by
\begin{align}
	H(\mathbf{k})=\sum^{5}_{a=1} d_a(\mathbf{k})\Gamma^{a}+\sum^{5}_{a<b=1}d_{ab}(\mathbf{k})\Gamma^{ab}
\end{align}
where $d_1=t(1+2\cos(x)\cos(y))$, $d_2=\lambda_{\upsilon}$, $d_3=\lambda_R(1-\cos(x)\cos(y))$, $d_4=-\sqrt{3}\lambda_R\sin(x)\sin(y)$, $d_{12}=-2t\cos(x)\sin(y)$, $d_{15}=\lambda_{SO}(2\sin(2x)-4\sin(x)\cos(y))$, $d_{23}=-\lambda_R\cos(x)\sin(y)$ and $d_{24}=\sqrt{3}\lambda_R\sin(x)\cos(y)$ with $x=k_x/2$ and $y=\sqrt{3}k_y/2$. The Dirac matrices are $\Gamma^{(1,2,3,4,5)}=(\tau_x\sigma_0,\tau_z \sigma_0,\tau_y\sigma_x,\tau_y\sigma_y,\tau_y\sigma_z)$, and $\Gamma^{ab}=[\Gamma^a,\Gamma^b]/(2i)$, where the Pauli matrices $\tau_k (\sigma_k)$ represent the sublattice (spin) degree of freedom. Note that the term $d_2\Gamma^2$ breaks inversion symmetry and induces a topological phase transition.

Fig. 1(a1,b1) shows the optical bounds in the nontrivial and trivial phases, respectively. As shown in Fig. 1(c1), the optical weight with $m=2t$ (and $3t$) saturates to a value greater than two after a certain frequency cutoff. This indicates that the phase is nontrivial. Because Kane-Mele model hosts $C_3$ rotation symmetry, the optical bound decays very slowly after the transition to the trivial phase. Fortunately, the decay can be significantly enhanced by increasing the optical gap, which is controlled by $\lambda_{SO}$ (see Appendix). The optical weight and quantum volume cannot represent the $Z_{2}$ invariant directly. Instead, they reflect the $Z_2$ topological signature related to bulk-boundary correspondence \cite{Prodan16}, which describes the relationship between the bulk topological invariant of a gapped system and the number of topologically protected boundary states. As shown in Fig. 1(d1), the double quantum volume consistently exceeds two in the nontrivial region, it suggests that the double quantum volume can be interpreted as the tightly upper bound on the number of boundary states. If the Hamiltonian preserves the symmetry of the lattice, $\overline{C}$ will be slightly larger than two, as is the case for TIs. The Rashba term, which breaks the $M_z$ mirror symmetry, serves a gap-closing role and mixes spin up and down states, which can reduce the value of $\overline{C}$ (see Fig. 3(b1) in Appendix). Note that the $C_6$ rotation symmetry can also lead to a slow decay of the optical bound upon transition to the trivial phase, because $g_{xx}(\mathbf{k})\ne g_{yy}(\mathbf{k})$. However, if $C_4$ rotation symmetry is present, we will have $g_{xx}(\mathbf{k})=g_{yy}(\mathbf{k})$, as we will demonstrate in the cases of TCI and HOTI below.


Next, we consider the TCI model of mirror-protected insulator \cite{Wieder20}:
\begin{align}
	H(\mathbf{k})&=t_1\tau_z\sigma_0[\cos(k_x)+\cos(k_y)]  \nonumber\\
	&+t_2\tau_x\sigma_0[\cos(k_x)-\cos(k_y)]  \nonumber\\
	&+t_{PH}\tau_0\sigma_0[\cos(k_x)+\cos(k_y)]  \nonumber\\
	&+\upsilon_{s}\tau_y\sigma_z\sin(k_x)\sin(k_y)+\upsilon_{m}\tau_z\sigma_0
\end{align}
where $\tau (\sigma)$ corresponds to the s and d-orbital (spin) degrees of freedom, it is invariant under the symmetries of the layer group $p4/mmm1'$ (see \cite{Wieder20}'s Table 1). Due to the presence of the mirror symmetry $M_z$, the topology of the TCI model is characterized by the mirror Chern number $C_{M_z}=2$ \cite{Wieder20}.

The optical bounds of the TCI model are shown in Fig. 1 (a2, b2). Furthermore, the optical weight rapidly saturates to a value greater than four as the frequency cutoff increases in the nontrivial phase, while the optical weight remains low in the trivial phase. Notably, as shown in Fig. 1(d2), both the double quantum volume and "pseudo-Chern number" are quantized and take the same value: four. This is an empirical observation under $C_4$ rotation symmetry with even-parity matrix elements. Due to the presence of $C_4$ rotation symmetry, $\mathcal{K}$, $\mathcal{V}$, and $\overline{C}$ rapidly decay and reach nearly the same value in the trivial and nontrivial phases, respectively, reflecting the near overlap between the optical bound and the generalized optical Hall conductivity. This indicates that the $Z_2$ topological signature can be clearly identified in this kind of systems.

Finally, we consider the HOTI model of quadrupole  insulator \cite{Benalcazar17,Benalcazar17b}:
\begin{align}
	H(\mathbf{k})&=[\gamma_x+\lambda_x\cos(k_x)]\Gamma_4+\lambda_x\sin(k_x)\Gamma_3  \nonumber\\
    &+[\gamma_y+\lambda_y\cos(k_y)]\Gamma_2+\lambda_y\sin(k_y)\Gamma_1,
\end{align}
where $\Gamma_0=\tau_3\tau_0$, $\Gamma_k=-\tau_2\tau_k(k=1,2,3)$, $\Gamma_4=\tau_1\tau_0$, $\tau_{1,2,3}$ represent the Pauli matrices, while $\tau_0$ is the $2 \times 2$ identity matrix. The model exhibits a quadrupole phase, $q_{xy} = 1/2$, with helical corner states when the parameters satisfy $\gamma_x/\lambda_x \in (-1, 1)$ and $\gamma_y/\lambda_y\in(-1,1)$. 

As shown in Fig. 2 (a1,a2), unlike the case of TCI, the optical bound and the generalized optical Hall conductivity are well separated in the nontrivial phase. This feature reflects on the larger difference between $\mathcal{K}$ and $\overline{C}$. The $Z_2$ topological signature is shown in Fig. 2 (c1,d1,c2,d2): the optical weight and quantum weight are greater than the number of boundary states, which equals two for $m\in(-1,1)$. Fig. 2 (d1) shows both $\mathcal{K}$ and $2\mathcal{V}$ exhibit discontinuous behavior \cite{Venuti07,Zanardi07,Carollo20} and nearly overlap, reflecting a topological phase transition from $\mathbf{p}^{\nu}=(1/2,1/2)$ to $\mathbf{p}^{\nu}=(0,0)$, characterized by band gap closing. In contrast, as shown in Fig. 2 (d2), a different type of topological phase transition from $\mathbf{p}^{\nu}=(1/2,1/2)$ to $\mathbf{p}^{\nu}=(0,1/2)$ occurs without gap closing, where both $\mathcal{K}$ and $2\mathcal{V}$ display continuous behavior and nearly overlap. Note that in the phase with $\mathbf{p}^{\nu}=(0,1/2)$, there is only one boundary state at the corners. As shown in Fig.2 (d1,d2), $\overline{C}$ is less than two, indicating that the Kubo form of MBC cannot fully capture the quadrupole phase. This demonstrates the limits of the Kubo form of MBC approach in characterizing the topology of HOTIs. Although the $Z_2$ topological signature cannot always be manifested by the Kubo form of MBC in HOTI systems, as shown in Fig. 2(e1, e2), the recovery of a Kubo form of MBC greater than two by adding a term that breaks mirror symmetry \cite{Li20} implies that the tight topological lower bound still exists through a specific topological deformation. As a result, the optical bound still provides the $Z_2$ topological signature, i.e., an odd number of pair boundary states. Note that recovering the tight topological lower bound upon inclusion of the broken-mirror-symmetry term does not constitute a general proof that such a bound exists.

\textit{Discussions}.---The refined TDI is particularly useful for systems that host completely flat bands, such as fractional quantum spin Hall insulator (FQSHI) \cite{Repellin14,Simon15,Liu16,Wu24b,Kang24,Kang25}. It is known that when the TDI is saturated, the FCI phase can be realized \cite{Roy14,Xie21}. In TRS systems, when the TDI is saturated, i.e., $\mathcal{K}=2\mathcal{V}=\overline{\overline{C}}=|C_+|+|C_-|$ \cite{Niu85,Onishi24,degeneracy}, we expect that the FQSHI phase can emerge. This suggests that "pseudo-Chern number" can be quantized. Since the TDI can also be applied to systems with broken TRS, in this case, we have $\mathrm{tr} g({\bf k})=2\sqrt{\det(g({\bf k}))}=\overline{\Omega}_{ab}({\bf k})=|\Omega_{ab}({\bf k})|$. This indicates that all components of the Berry curvature will be semi-positive (semi-negative) or positive (negative) in the FCI phases \cite{Wang21b,Wang23}. In other words, the imaginary part of the optical Hall conductivity will also be semi-positive (semi-negative) or positive (negative). 

In some cases, the optical weight is larger than the number of boundary states in the topologically trivial region near a topological phase transition. This makes the topology hard to identify. However, if one can experimentally tune a band‑inversion parameter to drive the transition, then, by a series of optical measurements, the topological transition point can be identified and the topologically nontrivial and trivial regions (i.e., the topological phase diagram) can be distinguished.

\textit{Conclusions}.---In summary, by focusing on two-dimensional TRS insulators, we have established a solid foundation and provided numerical demonstrations of how $Z_2$ topological signature can be identified through measurements of longitudinal optical responses. The optical weight and quantum volume sets an upper limit on the number of boundary states. The close relationship between geometry and topology, as revealed by the integrated TDI, is experimentally detectable. These concepts can also be applied to systems that break time-reversal symmetry, particularly in Moir\'e materials hosting flat Chern bands \cite{Kang24}. Our work opens a new avenue for probing the topology and quantized object in quantum geometry through measurements of optical responses. In future investigations, it will be important to rule out whether other factors—such as disorder \cite{Alisultanov24}, charge-density waves, spin-density waves, and many-body effects \cite{Basov11}—may cause a violation of the "topological inequality", and experimentally control the decay of the optical weight in the topologically trivial region.


Note added: After the submission, the author became aware of two works, Refs. \cite{Yu25, Jankowski25b}, which study the $Z_2$ bound using the Wilson loop and the quantum geometric bounds using the spin projection operator, respectively, for time-reversal symmetric insulators.

\textit{Acknowledgments}
The author thanks Hsiu-Chuan Hsu for providing the cluster for parallel numerical computation. The author is grateful to Ming-Che Chang for reading the manuscript and providing insightful comments. The author thanks Robert-Jan Slager for communicating his unpublished work. The author also thanks Po-Yao Chang, Hsiu-Chuan Hsu, and Jhih-Shih You for useful discussions. OpenAI GPT-5.6 and Gemini 3.6 were used to assist in deriving the proof of the upper and lower bound for the projection form of the maximal Berry curvature. All related mathematical derivations were verified and numerically confirmed by the author. This work was supported by the postdoctoral fellowship of the Ministry of Science and Technology (MOST) in Taiwan under Grants No. 110-2636-M-007-007 and No. 111-2811-M-004-001.


\bibliography{ref}

\clearpage
\appendix
{\centering\bfseries\large Appendix\par}
\vspace{4ex} 
\twocolumngrid 
\setcounter{secnumdepth}{1}

\section{Factors Influencing the Decay Behavior of Quantum Weight: Crystalline Symmetry, Optical Gap, and Inverse Mass Tensor}
Here, we only consider point group symmetry. To begin, we first examine rotational symmetry. As the decay behavior of the quantum weight near the transition from the topological non-trivial to trivial phase is strongly constrained by crystalline symmetries, we classify these crystalline symmetries into three groups, which we denote as $\mathcal{S}_1$, $\mathcal{S}_2$, and $\mathcal{S}_3$: (i) $\mathcal{S}_1$ contains $C_3$ and $C_6$ rotation symmetries, (ii) $\mathcal{S}_2$ contains $C_4$ rotation symmetry, and (iii) $\mathcal{S}_3$ includes crystalline symmetries that lack $C_3$, $C_4$ and $C_6$ rotation symmetries, as well as other crystalline symmetries with similar representation matrices. For other crystalline symmetries with a similar form of representation matrices for $C_3$ (i.e., $\left(\begin{smallmatrix}a&b\\ c&d\end{smallmatrix}\right),a,b,c,d\ne0$) and $C_4$ (i.e., $\left(\begin{smallmatrix}0&\pm1\\ \pm1&0\end{smallmatrix}\right)$), they will belong to $\mathcal{S}_1$  and $\mathcal{S}_2$, respectively.

In the following, we will see that the $C_3$ or $C_6$ rotation symmetry makes the quantum weight unequal, leading to $K_{xx}\ne K_{yy}$ , where $K_{ab}\equiv2\pi\int\frac{d^2k}{(2\pi)^2}g_{ab}({\bf k})$. In contrast, symmetries belonging to the group $\mathcal{S}_3$ have a relatively small effect, with no relation between $K_{xx}$ and $K_{yy}$. Only the $C_4$ rotation symmetry can yield $K_{xx}=K_{yy}$. Other composite symmetries with zero diagonal elements and $\pm1$ off-diagonal elements can also lead to $K_{xx}=K_{yy}$. If $K_{xx}\ne K_{yy}$ or in the presence of  $C_3(C_6)$ rotation symmetry, the decay of the quantum weight near the transition from the topological non-trivial to trivial phase is slow, making the optical conductivity difficult to use for identifying the topology. Fortunately, a larger optical gap $E_g$ (and/or effective mass) \cite{Onishi24,supp} (see Fig. 3) or flattening the band can cause $K_{xx}$ and $K_{yy}$ to decay rapidly.

To see how rotation symmetry provides a connection between elements of the quantum metric (matrix), let $R$ and $\mathcal{R}$ be point-group symmetry operations in real (momentum) and orbital bases, respectively: $x’_a=\sum_bR_{ab}x_b$ and $k’_a=\sum_bR^{-1}_{ab}k_b$. Additionally, $H’(\mathbf{k})=\mathcal{R}H(\mathbf{k’})\mathcal{R}^{-1}$ and $|u’_n(\mathbf{k})\rangle=\mathcal{R}|u_n(\mathbf{k’})\rangle$. Hence, the transformation of the quantum metric \cite{Ahn20} can be written as
\begin{align}
	g_{ba}(\mathbf{k})
	&=\sum_{b’a’}R_{bb’}R_{aa’}g_{b’a’}(\mathbf{k’}).
\end{align}
When substituting $C_3=\left(\begin{smallmatrix}-1/2&-\sqrt{3}/2\\ \sqrt{3}/2&-1/2\end{smallmatrix}\right)$ into Eq. (A1), we obtain
$g_{xx}(\mathbf{k})=\frac{1}{4}g_{xx}(\mathbf{k’})+\frac{\sqrt{3}}{4}g_{xy}(\mathbf{k’})+\frac{\sqrt{3}}{4}g_{yx}(\mathbf{k’})+\frac{3}{4}g_{y}(\mathbf{k’})$. Obviously, $g_{xx}(\mathbf{k})\ne g_{yy}(\mathbf{k’})$ unless $g_{xy}(\mathbf{k’})=0$. However, the $C_4=\left(\begin{smallmatrix}0&-1\\1&0\end{smallmatrix}\right)$ rotation symmetry can enforce $g_{xx}(\mathbf{k})=g_{yy}(\mathbf{k’})$, which lowers the "geometric effect" of the dispersive bands, as we will discuss below. As a result, we obtain $K_{xx}=K_{yy}$ for systems preserving $C_4$ rotation symmetry. The condition of $K_{xx}=K_{yy}$ is important for the rapid decay of the quantum weight near the transition from the topological nontrivial to trivial phase. In addition to $C_4$ rotation symmetry, the magnitude of the band-inversion parameter is also a key factor influencing the rapid decay of the quantum weight in certain cases (for example, spin-orbit coupling should be smaller than the nearest neighbor hopping in the Bernevig-Hughes-Zhang model \cite{Bernevig06,Ahn17}).

Another factor that influences the decay behavior of quantum weight is the optical gap ($E_g$) \cite{Onishi24} and the inverse mass tensor in the "topological inequality" \cite{Onishi24,supp}: $\frac{\pi e^2}{2}\frac{\mathcal{M}}{E_g}\ge\mathcal{K}\ge2\mathcal{V}\ge\overline{\overline{C}}\ge\sum_{s=1}^{n_t}|\mathcal{N}_{s}|$,
where $\mathcal{M}=\sum_{m,n,a}\int \frac{d\mathbf{k}}{(2\pi)^2}[M^{-1}_{mn}(\mathbf{k})]_{aa}\langle c^{\dagger}_{\mathbf{k}m}c_{\mathbf{k}n}\rangle$ and  $[M^{-1}_{mn}(\mathbf{k})]_{aa}=\sum_{\alpha\beta}U^{\dagger}_{m,\alpha}(\mathbf{k})\frac{\partial^2H_{\alpha\beta}(\mathbf{k})}{\partial(\hbar k_a)^2}U_{\beta,n}(\mathbf{k})$ is the inverse mass tensor \cite{Kubo57,Hazra19}. A larger value of $E_g$ can bring $\mathcal{K}$, $2\mathcal{V}$ and $\overline{\overline{C}}$ closer together. For example, in the Kane-Mele model, we can obtain a sufficiently large $E_g$ by setting a larger $\lambda_{SO}$, say $0.6t$ (see Fig. 3(b3)). The non-vanishing of quantum weight in the trivial phase can be interpreted as the "geometric effect", which is related to the inverse mass tensor. Therefore, flattening the band can also enhance the decay of the quantum weight in the trivial phase.

The condition for obtaining quantized quantum volume is subtle. Here, we list two classes of possible conditions: (i) the system preserves $C_4$ symmetry and has even-function matrix elements in the Hamiltonian, such as the TCI model of mirror-protected insulator \cite{Wieder20}; and (ii) the system has ideal flat bands \cite{Jie21,Chiu25}.

\begin{figure}[]
	\includegraphics[width=0.49\textwidth]{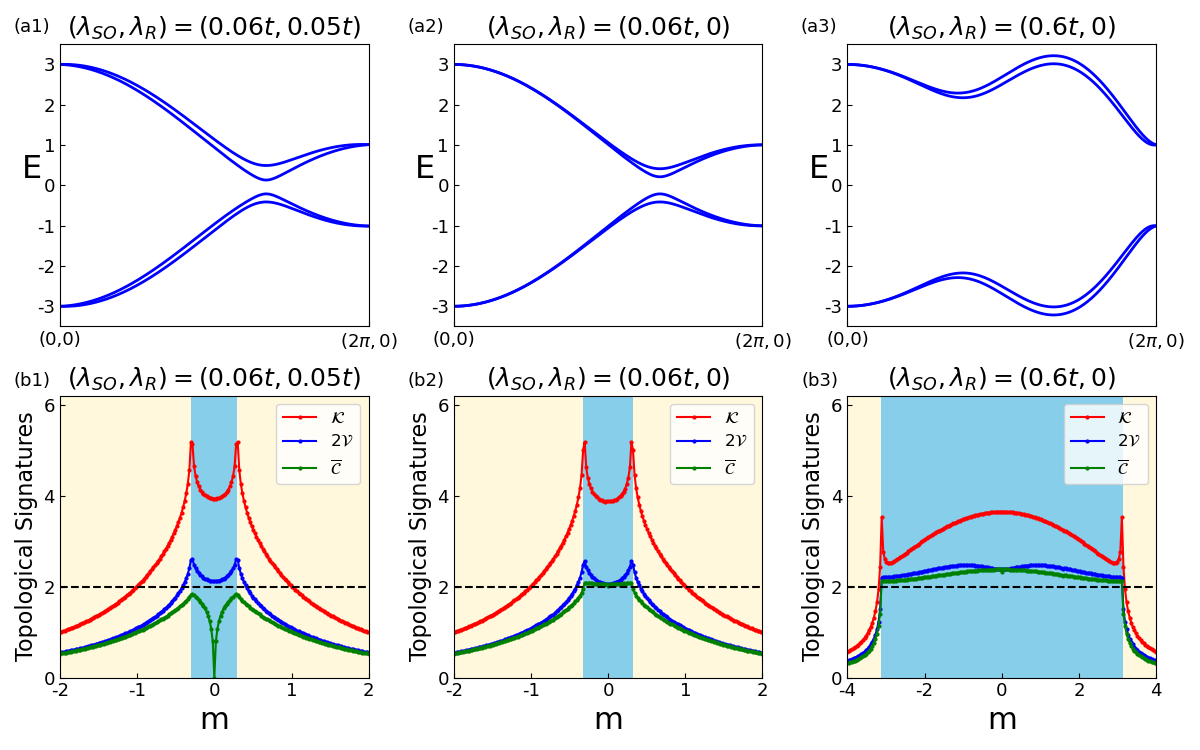}
	\caption{Band structure and topological signatures of the Kane-Mele model with three sets of $(\lambda_{SO},\lambda_{R})$ parameters. (a1-a3) Band structure with (a1) $(\lambda_{SO},\lambda_{R})=(0.06t,0.05t)$, (a2) $(\lambda_{SO},\lambda_{R})=(0.06t,0)$, and (a3) $(\lambda_{SO},\lambda_{R})=(0.6t,0)$, in which we set $\lambda_{\upsilon}=0.1t$. (b1-b3) The topological signatures $\mathcal{K}$, $2\mathcal{V}$ and $\overline{C}$  with (b1) $(\lambda_{SO},\lambda_{R})=(0.06t,0.05t)$, (b2) $(\lambda_{SO},\lambda_{R})=(0.06t,0)$, and (b3) $(\lambda_{SO},\lambda_{R})=(0.6t,0)$ as a function of $m$. The black dashed line represents the number of boundary states. Here we set $t=1$ and $\lambda_{\upsilon}=0.1t$. The presence of Rashba spin-orbit coupling, which breaks the mirror symmetry ($M_z$), will reduce the value of MBC.
	}
\end{figure}

\section{Tight Topological Lower Bound of Kubo form and Projection form Maximal Berry Curvature}

The Kubo form of the Berry curvature can reveal the effects of broken crystalline symmetry and visualize topological deformation. However, it possesses a topological lower bound only under certain conditions. This arises from a cancelled term, i.e., $-2\Im(r^{a}_{mn}(\bm{k})r^{b}_{nm}(\bm{k}))$ where $m,n\in \text{occ}$, in the expression for the Berry curvature of an individual band. The sum over all such cancelled terms vanishes, i.e., $\sum_{m\ne n}^{ \text{occ}}(-2\Im(r^{a}_{mn}(\bm{k})r^{b}_{nm}(\bm{k})))=0$. We demonstrate the conditions as follows.

Without loss of generality, we consider only four-band models. Our analysis can easily be generalized to systems with more than four bands. First, we write the expressions for the Abelian Berry curvature of the two individual occupied bands and the non-Abelian Berry curvature of the two occupied bands:
\begin{align}
	\Omega^0_{xy}(\bm{k})&=\Omega^{01}_{xy}(\bm{k})+F_0(\bm{k})  \\ \Omega^1_{xy}(\bm{k})&=\Omega^{10}_{xy}(\bm{k})+F_1(\bm{k})  \\
	\Omega_{xy}(\bm{k})&=F_0(\bm{k})+F_1(\bm{k})  
\end{align}
where $\Omega^{01}_{xy}(\bm{k})=-2\Im(r^{x}_{01}(\bm{k})r^{y}_{10}(\bm{k}))=-\Omega^{10}_{xy}$ and $F_n(\bm{k})=\Omega^{n2}_{xy}(\bm{k})+\Omega^{n3}_{xy}(\bm{k})=-2\Im(r^{x}_{n2}(\bm{k})r^{y}_{2n}(\bm{k}))-2\Im(r^{x}_{n3}(\bm{k})r^{y}_{3n}(\bm{k}))$ with $n=0,1$. We can see that $\overline{\Omega}_{xy}(\bm{k})\ge|\Omega^0_{xy}(\bm{k})|+|\Omega^1_{xy}(\bm{k})|$ when the cancelled term vanishes. However, our computation for HOTI shows that even with $\Omega^{01}_{xy}(\bm{k})$ non-vanishing, the inequality still holds (see Fig. 4). Hence, taking momentum integration of both sides of the inequality yields a tight topological lower bound in breaking TRS systems. As shown in Fig. 4, the cancelled terms and the remainder terms exhibit different behaviors. The cancelled terms have opposite signs and can vanish over the Brillouin zone. In the Kane-Mele model, the Rashba term breaks the mirror symmetry, which causes non-vanishing $\Omega^{01}_{xy}(\bm{k})$ and $\Omega^{10}_{xy}(\bm{k})$. Similarly, in HOTI, the broken mirror symmetry term \cite{Li20} also causes non-vanishing $\Omega^{01}_{xy}(\bm{k})$ and $\Omega^{10}_{xy}(\bm{k})$ and enlarges the remainder terms in the meantime. In other words, the topological deformation is visualized in the same phase.

\begin{figure}[]
	\includegraphics[width=0.49\textwidth]{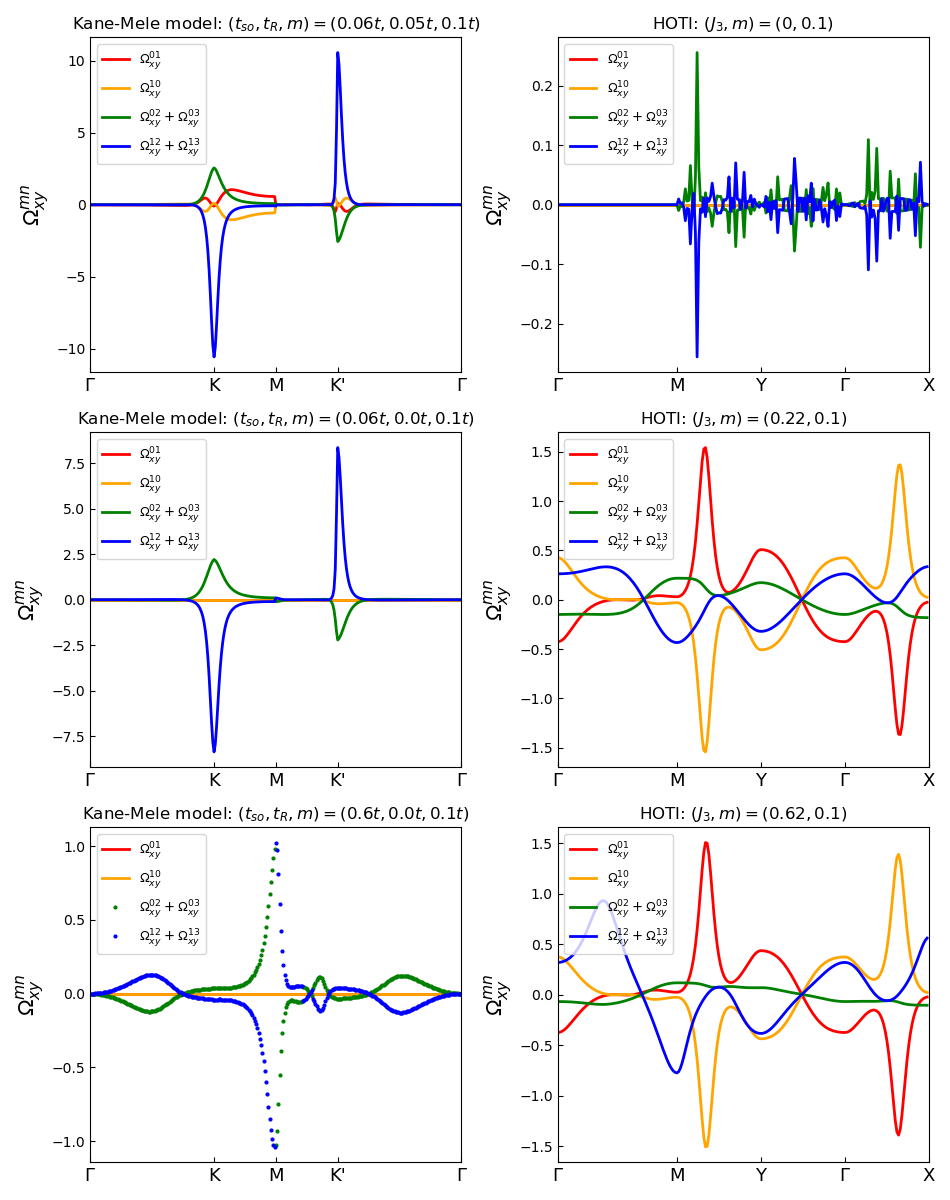}
	\caption{The components of the Berry curvature of the Kane-Mele model (first column) and HOTI (second column) along the path between high-symmetry points on the Brillouin zone. We can see that the components $\Omega^{01}_{xy}(\bm{k})$ and $\Omega^{10}_{xy}(\bm{k})$ have opposite signs and can vanish over the Brillouin zone. In contrast, $\Omega^{02}_{xy}(\bm{k})+\Omega^{03}_{xy}(\bm{k})$ and $\Omega^{12}_{xy}(\bm{k})+\Omega^{13}_{xy}(\bm{k})$ may or may not have opposite signs and are always non-vanishing. The remaining parameters are the same as in Fig. 1 and Fig. 2. Here we denote $J_3$ as the magnitude of the broken mirror term \cite{Li20}.}
\end{figure}

\begin{figure}[]
	\includegraphics[width=0.49\textwidth]{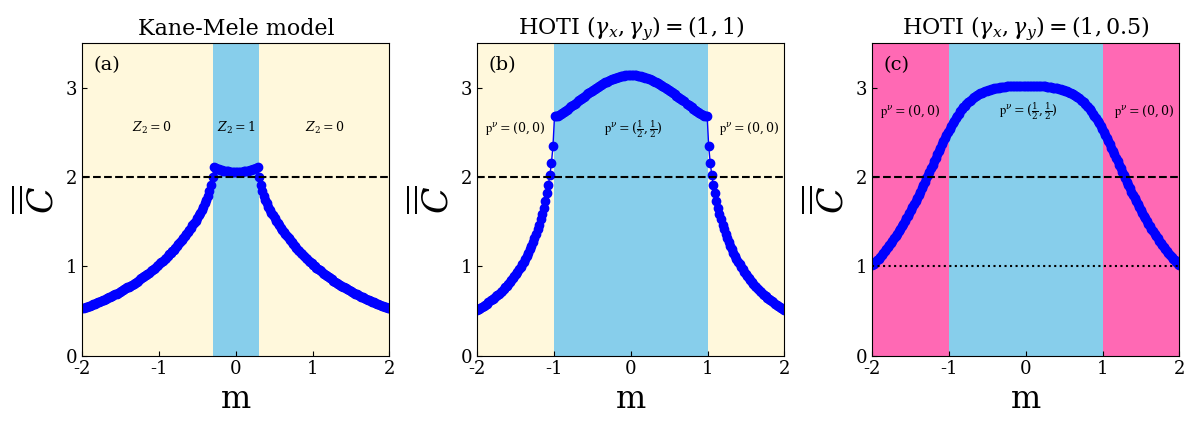}
	\caption{The "pseudo-Chern number" of the projection form of MBC, $\overline{\overline{C}}$, as a function of $m$ for the Kane-Mele model (a) and HOTI (b,c). The black dashed and dotted lines represent the number of boundary states. The remaining parameters are the same as in Fig. 1 and Fig. 2.}
\end{figure}

Unlike the Kubo form of MBC, the projection form of MBC is not sensitive to broken crystalline symmetry and always manifests a tight topological bound, for which we provide a proof in the supplemental material. Fig. 5 shows that the "pseudo-Chern number" of the projection form of MBC is always larger than the tight topological bound, which is two in the Kane-Mele model and two or one in HOTI. In contrast, the Kubo form of MBC's "pseudo-Chern number" fails to give the tight topological lower bound when certain crystalline symmetries are broken.

\clearpage           

\thispagestyle{empty}
\mbox{}
\pagebreak
\newpage
\onecolumngrid
\begin{center}
	\textbf{\large Supplemental Material for "$Z_2$ topological signatures of the optical bound on maximal Berry curvature: Application to two-dimensional time-reversal symmetric insulators"}
\end{center}

\author{Pok-Man Chiu}\email{pokman2011@gmail.com}
\affiliation{Graduate Institute of Applied Physics, National Chengchi University, Taipei City 11605, Taiwan}
\affiliation{Department of Physics, National Tsing Hua University, Hsinchu 30013, Taiwan}

\setcounter{equation}{0}
\setcounter{figure}{0}
\setcounter{table}{0}
\setcounter{page}{1}
\makeatletter
\renewcommand{\theequation}{S\arabic{equation}}
\renewcommand{\thefigure}{S\arabic{figure}}
\renewcommand{\bibnumfmt}[1]{[S#1]}
\renewcommand{\citenumfont}[1]{S#1}

\onecolumngrid

In this supplementary material, we review several aspects of quantum geometry and provide a detailed derivation of the refined TDI, the optical bound, the SWM sum rule, and the relationship between the optical bound and quantum weight. Finally, we present additional numerical results to enhance the understanding of the main concepts.

\subsection{1: Quantum geometry of occupied bands}
In this section, we review some basic concepts of quantum geometry. We follow the approach recently introduced by Ahn et al. \cite{Ahn20,Ahn22}, who utilize an analogous definition from differential geometry \cite{Chern00} to derive the quantum metric tensor and Berry curvature, which are equivalent to the conventional approach \cite{Provost80,Marzari97}. A key step in their method involves defining the tangent vector basis and the associated inner product for quantum states, which can be treated as a manifold. In the context of optical transitions between states $m$ and $n$ with $m\neq n$, one can choose the tangent vector as follows:
\begin{equation}
	\hat{e}^{mn}_{a}(\mathbf{k}) \equiv r^{a}_{mn}(\mathbf{k}) | u_{m}(\mathbf{k}) \rangle \langle u_{n}(\mathbf{k}) |, \quad a=1,...,d.
\end{equation}
where the interband Berry connection is given by $r^a_{mn}(\mathbf{k})=i\langle u_m(\mathbf{k})|\partial_a|u_n(\mathbf{k})\rangle$. For the complex Riemannian structure, one can use the Hilbert-Schmidt inner product:
\begin{equation}
	(A,B)=\sum_{m,n}A^{\ast}_{mn}B_{mn}=\mathrm{tr}(A^{\dagger}B).
\end{equation}
From the two definitions above, we can obtain the Hermitian metric tensor for any pair of bands,
\begin{equation}
	Q^{mn}_{ba}(\mathbf{k})\equiv(\hat{e}^{mn}_{b}(\mathbf{k}),\hat{e}^{mn}_{a}(\mathbf{k}))=r^{b}_{nm}(\mathbf{k})r^{a}_{mn}(\mathbf{k}).
\end{equation}
This interband-resolved geometric quantity, $Q^{mn}_{ba}(\mathbf{k})=g^{mn}_{ba}(\mathbf{k}) - i\Omega^{mn}_{ba}(\mathbf{k})/2$ is the component of quantum geometric tensor \cite{Marzari97} in multiband case. Note that there is no band summation. $Q^{mn}_{ba}(\mathbf{k})$ itself can be understood as the quantum geometric tensor for manifold of two-band subspace. For the occupied bands case, one can generalized $\hat{e}^{mn}_{a}(\mathbf{k})$ to
\begin{equation}
	\hat{e}_{a}(\mathbf{k}) \equiv \sum_{n\in occ}\sum_{m\in unocc} | u_{m}(\mathbf{k}) \rangle r^{a}_{mn}(\mathbf{k}) \langle u_{n}(\mathbf{k}) |, 
\end{equation}
as the tangent vector in complex Grassmannian manifold of occupied bands \cite{Ahn22}. Hence we obtain the quantum geometric tensor of the occupied bands:
\begin{equation}
	Q_{ba}(\mathbf{k})\equiv(\hat{e}_{b}(\mathbf{k}),\hat{e}_{a}(\mathbf{k}))=\sum_{n\in occ}\sum_{m\in unocc}r^{b}_{nm}(\mathbf{k})r^{a}_{mn}(\mathbf{k}).
\end{equation}
Further, the quantum geometric tensor of the occupied bands can be expressed as \cite{Ahn20},
\begin{align}
	Q_{ab}(\mathbf{k})=\sum_{n\in{occ}}\sum_{m\in{unocc}}r^{a}_{nm}(\mathbf{k})r^{b}_{mn} (\mathbf{k}) 
	=\sum_{n\in{occ}}(g^{nn}_{ab}(\mathbf{k})-\frac{i}{2}\Omega^{nn}_{ab}(\mathbf{k}))
	=g_{ab}(\mathbf{k})-\frac{i}{2}\Omega_{ab}(\mathbf{k}),
\end{align}
where the non-Abelian quantum metric and non-Abelian Berry curvature are
\begin{align}
	Q_{ab}(\mathbf{k})&=\sum_{n\in occ}\sum_{m\in{unocc}}\langle \partial_{a}u_{n}(\mathbf{k})|u_m(\mathbf{k})\rangle\langle u_m(\mathbf{k})|\partial_b|u_{n}(\mathbf{k})\rangle  \nonumber\\
	&=\sum_{n\in occ}\langle \partial_{a}u_{n}(\mathbf{k})|(1-P(\mathbf{k}))|\partial_bu_{n}(\mathbf{k})\rangle  \nonumber\\
	&=\sum_{n\in occ}[g^{n,n}_{ab}(\mathbf{k})-\frac{i}{2}\Omega^{nn}_{ab}(\mathbf{k})],
\end{align}

Here we use a relation \cite{Ahn22}
\begin{align}
	\sum_{m\in{unocc}}r^{a}_{n_{1}m}(\mathbf{k})r^{b}_{mn_{2}}(\mathbf{k})&=\sum_{m\in{unocc}}\langle \partial_{a}u_{n_1}(\mathbf{k})|u_m(\mathbf{k})\rangle\langle u_m(\mathbf{k})|\partial_b|u_{n_2}(\mathbf{k})\rangle  \nonumber\\
	&=\langle \partial_{a}u_{n_1}(\mathbf{k})|(1-P(\mathbf{k}))|\partial_bu_{n_2}(\mathbf{k})\rangle  \nonumber\\
	&=g^{n_{1}n_{2}}_{ab}(\mathbf{k})-\frac{i}{2}\Omega^{n_{1}n_{2}}_{ab}(\mathbf{k}),
\end{align}
where the projection operator is give by $P(\mathbf{k})=\sum_{n\in{occ}}|u_n(\mathbf{k}) \rangle \langle u_n(\mathbf{k})|$. 

One also can express the quantum geometric tensor using the projection operator \cite{Peotta15,Jonah22}, with a form $P(\mathbf{k})=U(\mathbf{k})U^{\dagger}(\mathbf{k})$, where $U(\mathbf{k})$ is the $N\times N_{occ}$ matrix whose columns are the eigenvectors of the occupied bands. Here $N$ is the number of band. Similarly, one can define a tangent vector, which is a covariant derivative acting on $U(\mathbf{k})$, i.e., $(1-U(\mathbf{k})U^{\dagger}(\mathbf{k}))\partial_aU(\mathbf{k})$. Using the Hilbert-Schmidt inner product again, the quantum geometric tensor can be defined as
\begin{align}
	Q_{ab}(\mathbf{k})=\mathrm{tr}(((1-U(\mathbf{k})U^{\dagger}(\mathbf{k}))\partial_aU(\mathbf{k}))^{\dagger}(1-U(\mathbf{k})U^{\dagger}(\mathbf{k}))\partial_aU(\mathbf{k})).
\end{align}
After some algebra (see Ref. \cite{Jonah22}), we obtain:
\begin{align}
	Q_{ab}(\mathbf{k})=\mathrm{tr}(P(\mathbf{k})\partial_a P(\mathbf{k}) \partial_b P(\mathbf{k}))=g_{ab}(\mathbf{k})-\frac{i}{2}\Omega_{ab}(\mathbf{k}),
\end{align} 
where the non-Abelian quantum metric and non-Abelian Berry curvature are
\begin{align}
	g_{ab}(\mathbf{k})=\frac{1}{2}\mathrm{tr}(\partial_a P(\mathbf{k}) \partial_b P(\mathbf{k})),  \quad 
	\Omega_{ab}(\mathbf{k})=i\mathrm{tr}(P(\mathbf{k}) [\partial_a P(\mathbf{k}),\partial_b P(\mathbf{k})]).
\end{align}

\subsection{2. Refined trace-determinant inequality for TIs/TCIs with a tight topological lower bound}
In this subsection, we will prove the refined TDI for TIs/TCIs using the method of projection \cite{Prodan09,Prodan10,Sheng11,Li13,Lin24}. As enforced by TRS, we know that the Chern number of the occupied bands is zero in the usual basis \cite{Moore07}. However, when projecting the occupied bands onto different spin (pseudospin) or topological sectors, i.e., replacing $P(\mathbf{k})=\sum^{occ}_{n=1}|u_{n}(\mathbf{k}) \rangle \langle u_{n}(\mathbf{k})|$ by $P(\mathbf{k})\hat{s}P(\mathbf{k})$, we can obtain nonzero Chern number for each topological sector in a TRS system, where $\hat{s}$ is a suitable operator that can isolate the topology in different sectors. Crucially, the choice of $\hat{s}$ is subject to the following two constraints \cite{Kivelson82,Lin24}:
\begin{align}
	P(\mathbf{k})|\psi^{\pm}_i(\mathbf{k})\rangle&=|\psi^{\pm}_i(\mathbf{k})\rangle,  \\
	P(\mathbf{k})\hat{s}P(\mathbf{k})|\psi^{\pm}_i(\mathbf{k})\rangle&=\lambda^{\pm}_i|\psi^{\pm}_i(\mathbf{k})\rangle,
\end{align}
where the nonzero eigenvalues $\lambda^{\pm}_i$ correspond to distinct spin (pseudospin) sectors, which are labeled by $\pm$. We can view both $P(\mathbf{k})$ and $P(\mathbf{k})\hat{s}P(\mathbf{k})$ as flattening Hamiltonians. According to the spectral theorem \cite{Goldbring21} or the spectral representation \cite{Kreyszig78}, we can re-express $P(\mathbf{k})$ using the eigenstates $|\psi^\pm_i(\mathbf{k}) \rangle$ of $P(\mathbf{k})\hat{s}P(\mathbf{k})$, i.e., $P(\mathbf{k})=\sum^{N_+}_{i=1}|\psi^+_i(\mathbf{k}) \rangle \langle \psi^+_i(\mathbf{k})|+\sum^{N_-}_{i=1}|\psi^-_i(\mathbf{k}) \rangle \langle \psi^-_i(\mathbf{k})|$. As a result, we obtain $P(\mathbf{k})=\sum^{N_+}_{i=1}|\psi^+_i(\mathbf{k}) \rangle \langle \psi^+_i(\mathbf{k})|+\sum^{N_-}_{i=1}|\psi^-_i(\mathbf{k}) \rangle \langle \psi^-_i(\mathbf{k})|=\sum^{occ}_{n=1}|u_{n}(\mathbf{k}) \rangle \langle u_{n}(\mathbf{k})|$. In other words, $P(\mathbf{k})$ has two different sets of eigenstates $\{|\psi^\pm_i(\mathbf{k}) \rangle\}$
and $\{|u_{n}(\mathbf{k}) \rangle\}$, which represent the same system. In fact, $|\psi^\pm_i(\mathbf{k}) \rangle$ is simply a linear combination of $|u_{n}(\mathbf{k}) \rangle$. 


For the Kane-Mele model and TCI, we choose $\hat{s}=\tau_0\sigma_z$, where $\tau_0$ is the two-dimensional identity matrix. Note that the choice of the spin operator should match the basis of the Hamiltonian. As pointed out in Ref. \cite{Lin24}, only the lowest and topmost spin-resolved bands are physically relevant. We denote the projection of the spin-resolved band as $P_\pm(\mathbf{k})=|\psi^\pm(\mathbf{k})\rangle\langle \psi^\pm(\mathbf{k})|$. Using these two spin-resolved band projections, we obtain the corresponding non-Abelian quantum metric and non-Abelian Berry curvature: 
\begin{align}
	g'_{ab}(\mathbf{k})=\frac{1}{2}\mathrm{tr}(\partial_a P'(\mathbf{k}) \partial_b P'(\mathbf{k})), \quad \Omega'_{ab}(\mathbf{k})=i\mathrm{tr}(P'(\mathbf{k})[\partial_a P'(\mathbf{k}),\partial_b P'(\mathbf{k})]),
\end{align}
where $P'(\mathbf{k})=P_+(\mathbf{k})+P_-(\mathbf{k})$. Obviously, we have $g'_{ab}(\mathbf{k})=g_{ab}(\mathbf{k})$ since $|\psi^\pm_i(\mathbf{k}) \rangle$ and $|u_{n}(\mathbf{k}) \rangle$ represent the same projection operator. A crucial part of proving the refined TDI and obtaining the topological lower bound is the general inequality: $2\sqrt{\det g(\mathbf{k})}\ge\overline{\overline{\Omega}}_{ab}(\mathbf{k})\ge\sum^{\text{occ}}_{n=1}|\Omega^{n}_{ab}(\mathbf{k})|$, which we have proved below. On the other hand, by the sum rule of the Berry curvature or Chern number \cite{Avron83}, we have $\Omega'_{ab}(\mathbf{k})=\Omega^{+}_{ab}(\mathbf{k})+\Omega^{-}_{ab}(\mathbf{k})$. Hence, it follows that $\overline{\overline{\Omega'}}_{ab}(\mathbf{k})\ge|\Omega^{+}_{ab}(\mathbf{k})|+|\Omega^{-}_{ab}(\mathbf{k})|\ge\Omega'_{ab}(\mathbf{k})$. Note that before projecting onto different topological sectors, each occupied band does not have a well-defined Berry curvature. Finally, we obtain the TDI for TIs and TCIs:
\begin{align}
	g'_{xx}(\mathbf{k})+g'_{yy}(\mathbf{k})\ge2\sqrt{\det{g'(\mathbf{k})}}\ge\overline{\overline{\Omega'}}_{ab}(\mathbf{k})\ge|\Omega^{+}_{ab}(\mathbf{k})|+|\Omega^{-}_{ab}(\mathbf{k})|.
\end{align}
In other words, the quantum metric is the same for both the original and spin-resolved bands, but the bulk topology over the full Brillouin zone only manifests in the spin-resolved bands. After integrating over the momentum variable in the first Brillouin zone, we obtain the refined inequality for TIs and TCIs with a topological lower bound:
\begin{align}
	\int\frac{d^2\bf{k}}{2\pi}(g'_{xx}(\mathbf{k})+g'_{yy}(\mathbf{k}))\ge\int\frac{d^2\bf{k}}{2\pi}(2\sqrt{\det{g'(\mathbf{k})}})\ge\int\frac{d^2\bf{k}}{2\pi}\overline{\overline{\Omega'}}_{ab}(\mathbf{k})\ge \left|\int\frac{d^2\bf{k}}{2\pi}\Omega^{+}_{ab}(\mathbf{k})\right|+\left|\int\frac{d^2\bf{k}}{2\pi}\Omega^{-}_{ab}(\mathbf{k})\right|,
\end{align}
where we use the fact that $\int\frac{d^2\bf{k}}{2\pi}|\Omega^{\pm}_{ab}(\mathbf{k})|\ge\left|\int\frac{d^2\bf{k}}{2\pi}\Omega^{\pm}_{ab}(\mathbf{k})\right|$. We label it as $\mathcal{K'}\ge2\mathcal{V'}\ge\overline{\overline{C'}}\ge|C_{+}|+|C_{-}|$,
where $\mathcal{K'}$, $\mathcal{V'}$, $\overline{\overline{C'}}$, and $C_{\pm}$ are the quantum weight, quantum volume, "pseudo-Chern number", and Chern number of the spin-resolved bands, respectively. Our proof can be straightforwardly generalized to systems of more than four bands.

Finally, we will prove the crucial part of the refined TDI: $2\sqrt{\det(g(\mathbf{k}))}\ge\overline{\Omega}_{ab}(\mathbf{k})$ and $2\sqrt{\det g(\mathbf{k})}\ge\overline{\overline{\Omega}}_{ab}(\mathbf{k})\ge\sum^{\text{occ}}_{n=1}|\Omega^{n}_{ab}(\mathbf{k})|$. First, we prove the inequality $2\sqrt{\det(g(\mathbf{k}))}\ge\overline{\Omega}_{ab}(\mathbf{k})$. Using the Cauchy-Schwarz inequality \cite{Lax02}: $\langle \mathbf{u}|\mathbf{u}\rangle\langle \mathbf{v}|\mathbf{v}\rangle\ge|\langle \mathbf{u}|\mathbf{v}\rangle|^2$, and substituting $\mathbf{u}=| u_{m}(\mathbf{k}) \rangle r^{a}_{mn}(\mathbf{k}) \langle u_{n}(\mathbf{k}) |$ and $\mathbf{v}=| u_{m}(\mathbf{k}) \rangle r^{b}_{mn}(\mathbf{k}) \langle u_{n}(\mathbf{k}) |$ into it, we have $g^{mn}_{aa}(\mathbf{k})g^{mn}_{bb}(\mathbf{k})\ge (g^{mn}_{ab}(\mathbf{k}))^2+(\Omega^{mn}_{ab}(\mathbf{k})/2)^2$. By moving $(g^{mn}_{ab}(\mathbf{k}))^2$ to the left-hand side, taking the square root, and performing the band summation, we obtain
\begin{align}
	\sum_{n\in occ}\sum_{m\in unocc}\sqrt{g^{mn}_{aa}(\mathbf{k})g^{mn}_{bb}(\mathbf{k})- \left(g^{mn}_{ab}(\mathbf{k})\right)^2}\ge\frac{1}{2}\sum_{n\in occ}\sum_{m\in unocc}|\Omega^{mn}_{ab}(\mathbf{k})|.
\end{align}
On the other hand, we have
\begin{align}
	&\quad\sum_{n\in occ}\sum_{m\in unocc}\sqrt{g^{mn}_{aa}(\mathbf{k})g^{mn}_{bb}(\mathbf{k})- (g^{mn}_{ab}(\mathbf{k}))^2}  \nonumber\\
	&=\sum_{n\in occ}\sum_{m\in unocc}\sqrt{\left(\sqrt{g^{mn}_{aa}(\mathbf{k})g^{mn}_{bb}(\mathbf{k})}-g^{mn}_{ab}(\mathbf{k})\right)\left(\sqrt{g^{mn}_{aa}(\mathbf{k})g^{mn}_{bb}(\mathbf{k})}+g^{mn}_{ab}(\mathbf{k})\right)}  \nonumber\\
	&\le\sqrt{\sum_{n\in occ}\sum_{m\in unocc}(\sqrt{g^{mn}_{aa}(\mathbf{k})g^{mn}_{bb}(\mathbf{k})}-g^{mn}_{ab}(\mathbf{k}))}\sqrt{\sum_{n\in occ}\sum_{m\in unocc}(\sqrt{g^{mn}_{aa}(\mathbf{k})g^{mn}_{bb}(\mathbf{k})}+g^{mn}_{ab}(\mathbf{k}))}  \nonumber\\
	&=\sqrt{\left(\sum_{n\in occ}\sum_{m\in unocc}\sqrt{g^{mn}_{aa}(\mathbf{k})g^{mn}_{bb}(\mathbf{k})}\right)^2-\left(\sum_{n\in occ}\sum_{m\in unocc}g^{mn}_{ab}(\mathbf{k})\right)^2}  \nonumber\\
	&\le\sqrt{\left(\sum_{n\in occ}\sum_{m\in unocc}g^{mn}_{aa}(\mathbf{k})\right)\left(\sum_{n\in occ}\sum_{m\in unocc}g^{mn}_{bb}(\mathbf{k})\right)-\left(\sum_{n\in occ}\sum_{m\in unocc}g^{mn}_{ab}(\mathbf{k})\right)^2}  \nonumber\\
	&=\sqrt{\det{g}(\mathbf{k})}.  
\end{align}
In the above derivation, we use the summation form of the Cauchy-Schwarz inequality for the two inequality symbols \cite{Ozawa21}. Therefore, we obtain the desired inequality $2\sqrt{\det{g}(\mathbf{k})}\ge\overline{\Omega}_{xy}(\mathbf{k})$. Obviously, we have $\overline{\Omega}_{xy}(\mathbf{k})\ge|\Omega_{xy}(\mathbf{k})|$.

Now we prove the inequality $2\sqrt{\det g(\mathbf{k})}\ge\overline{\overline{\Omega}}_{ab}(\mathbf{k})\ge\sum^{\text{occ}}_{n=1}|\Omega^{n}_{ab}(\mathbf{k})|$. To begin, we define two operators $X(\mathbf{k})=P(\mathbf{k})\partial_aP(\mathbf{k})(1-P(\mathbf{k}))$ and $Y(\mathbf{k})=P(\mathbf{k})\partial_bP(\mathbf{k})(1-P(\mathbf{k}))$ to re-express the quantum metric and projection form of MBC as $g_{aa}(\mathbf{k})=\mathrm{tr}(X(\mathbf{k})X^{\dagger}(\mathbf{k}))=||X(\mathbf{k})||_{\text{HS}}$, $g_{ab}(\mathbf{k})=\Re[\mathrm{tr}(X(\mathbf{k})Y^{\dagger}(\mathbf{k}))]$, and $\overline{\overline{\Omega}}_{ab}(\mathbf{k})=\mathrm{tr}|X(\mathbf{k})Y^{\dagger}(\mathbf{k})-Y(\mathbf{k})X^{\dagger}(\mathbf{k})|=||X(\mathbf{k})Y^{\dagger}(\mathbf{k})-Y(\mathbf{k})X^{\dagger}(\mathbf{k})||_1$, where $||X||_{\text{HS}}$ and $||X||_1$ are the Hilbert–Schmidt norm and Schatten-1 norm, respectively. An important step is to separate the operator $Y(\mathbf{k})$ into parallel and perpendicular parts, i.e., $Y(\mathbf{k})=Y_{\perp}(\mathbf{k})+c(\mathbf{k})X(\mathbf{k})$, where $c(\mathbf{k})=\frac{\Re[\mathrm{tr}(X(\mathbf{k})Y^{\dagger}(\mathbf{k}))]}{||X(\mathbf{k})||^2_{\text{HS}}}$. Hence we have $\Re[\mathrm{tr}(X(\mathbf{k})Y_{\perp}^{\dagger}(\mathbf{k}))]=0$, and $X(\mathbf{k})Y^{\dagger}(\mathbf{k})-Y(\mathbf{k})X^{\dagger}(\mathbf{k})=X(\mathbf{k})Y_{\perp}^{\dagger}(\mathbf{k})-Y_{\perp}(\mathbf{k})X^{\dagger}(\mathbf{k})$. Here we will use the triangle inequality
\begin{align}
	||A||_1+||B||_1\ge||A+B||_1,
\end{align}
and the Schatten Hölder inequality
\begin{align}
	||A||_{\text{HS}}||B||_{\text{HS}}\ge||AB||_1.
\end{align}
By substituting $A=X(\mathbf{k})Y_{\perp}^{\dagger}(\mathbf{k})$ and $B=-Y_{\perp}(\mathbf{k})X^{\dagger}(\mathbf{k})$ into the triangle inequality, we have
\begin{align}
	||X(\mathbf{k})Y_{\perp}^{\dagger}(\mathbf{k})||_1+||-Y_{\perp}(\mathbf{k})X^{\dagger}(\mathbf{k})||_1\ge||X(\mathbf{k})Y_{\perp}^{\dagger}(\mathbf{k})-Y_{\perp}(\mathbf{k})X^{\dagger}(\mathbf{k})||_1
\end{align}
Similarly, substituting $A=X(\mathbf{k})$ and $B=Y_{\perp}^{\dagger}(\mathbf{k})$ nto the Schatten–Hölder inequality, the left-hand side of S(21) yields an upper bound
\begin{align}
	2||X(\mathbf{k})||_{\text{HS}}||Y_{\perp}(\mathbf{k})||_{\text{HS}}\ge||X(\mathbf{k})Y_{\perp}^{\dagger}(\mathbf{k})||_1.
\end{align}
Combining Eqs. S(21) and S(22), we have
\begin{align}
	2||X(\mathbf{k})||_{\text{HS}}||Y_{\perp}(\mathbf{k})||_{\text{HS}}\ge||X(\mathbf{k})Y_{\perp}^{\dagger}(\mathbf{k})-Y_{\perp}(\mathbf{k})X^{\dagger}(\mathbf{k})||_1.
\end{align}
Using the fact $||Y_{\perp}(\mathbf{k})||^2_{\text{HS}}=||Y(\mathbf{k})||^2_{\text{HS}}-\frac{(\Re\mathrm{tr}X(\mathbf{k})Y^{\dagger}(\mathbf{k}))^2}{||X(\mathbf{k})||^2_{\text{HS}}}$, it yields $||X(\mathbf{k})||^2_{\text{HS}}||Y_{\perp}(\mathbf{k})||^2_{\text{HS}}=\det{g(\mathbf{k})}$. In other words, we obtain
\begin{align}
	2\sqrt{\det{g(\mathbf{k})}}\ge\overline{\overline{\Omega}}_{ab}(\mathbf{k}).
\end{align}

To prove that $\overline{\overline{\Omega}}_{ab}(\mathbf{k})\ge\sum^{\text{occ}}_{n=1}|\Omega^{n}_{ab}(\mathbf{k})|$, it is equivalent to prove the following inequality:
\begin{align}
	\mathrm{tr}|A|\ge\sum_n{|A_{nn}|},
\end{align}
which can be obtained using the singular value decomposition, the triangle inequality, and the Cauchy–Schwarz inequality. We first take the singular value decomposition of $A$, i.e., $A=USV^{\dagger}$, where $U$ and $V$ are unitary matrices and $S=\text{diag}(s_1,s_2,\cdots,s_N)$ is the singular values. The Schatten 1-norm and the matrix elements of $A$ can then be written as $\mathrm{tr}|A|=\sum^N_{n=1}s_n$ and $A_{nn}=\sum^N_{n=1}s_nU_{in}V^{\ast}_{in}$. By applying the triangle inequality, we have
\begin{align}
	\sum^N_{n=1}|A_{nn}|\le\sum^N_{i=1}\sum^N_{n=1}s_n|U_{in}||V_{in}|=\sum^N_{n=1}s_n\left(\sum^N_{i=1}|U_{in}||V_{in}|\right).
\end{align}
Further, using the Cauchy–Schwarz inequality for the vectors $(|U_{1n}|,\cdots,|U_{Nn}|)$ and $(|V_{1n}|,\cdots,|V_{Nn}|)$, we have
\begin{align}
	\sum^N_{i=1}|U_{in}||V_{in}|\le\sqrt{\sum^N_{i=1}|U_{in}|^2}\sqrt{\sum^N_{i=1}|V_{in}|^2}.
\end{align}
Using the fact $\sum^N_{i=1}|U_{in}|^2=1=\sum^N_{i=1}|V_{in}|^2$, we finally obtain
\begin{align}
	\mathrm{tr}|A|\ge\sum_n{|A_{nn}|}.
\end{align}
In our case, we set $A=iP(\mathbf{k})[\partial_a P(\mathbf{k}),\partial_b P(\mathbf{k})]$. Therefore we have $A_{nn}=\langle u_{n}(\mathbf{k}) i[\partial_a P(\mathbf{k}), \partial_b P(\mathbf{k})]| u_{n}(\mathbf{k}) \rangle=\langle u_{n}(\mathbf{k}) i[\partial_a P_n(\mathbf{k}), \partial_b P_n(\mathbf{k})]| u_{n}(\mathbf{k}) \rangle=\Omega^{n}_{ab}(\mathbf{k})$ and $\mathrm{tr}|A|=\overline{\overline{\Omega}}_{ab}(\mathbf{k})$. Finally, we obtain
\begin{align}
	\overline{\overline{\Omega}}_{ab}(\mathbf{k})\ge\sum^{\text{occ}}_{n=1}|\Omega^{n}_{ab}(\mathbf{k})|.
\end{align}

\subsection{3: Optical bound}
In this subsection, we derive the optical bound from the refined TDI. Firstly, we explicitly write down the expressions for the non-Abelian quantum metric and the non-Abelian MBC.
\begin{align}
	&g_{ab}(\bm{k})=\sum_{m\in occ}\sum_{n\in unocc}\Re(r^{a}_{mn}(\bm{k})r^{b}_{nm}(\bm{k})),  \\
	&\overline{\Omega}_{ab}(\bm{k})=\sum_{m\in occ}\sum_{n\in unocc}|-2\Im(r^{a}_{mn}(\bm{k})r^{b}_{nm}(\bm{k}))|.
\end{align}
Then the refined TDI becomes
\begin{align}
	\sum_{m\in occ}\sum_{n\in unocc}  \Re(r^{x}_{mn}r^{x}_{nm}+r^{y}_{mn}r^{y}_{nm})  
	-\sum_{m\in occ}\sum_{n\in unocc}|-2\Im(r^{x}_{mn}r^{y}_{nm})|\ge0.
\end{align}
Here, we neglect the term $2\sqrt{\det{g}(\mathbf{k})}$. Using the identities $r^{a}_{mn}=\hbar\upsilon^{a}_{mn}/(iE_{mn})$, and $|\upsilon^{x}_{mn}\pm i\upsilon^{y}_{mn}|^{2}=|\upsilon^{x}_{mn}|^{2}+|\upsilon^{y}_{mn}|^{2}\mp i(\upsilon^{x}_{mn}\upsilon^{y}_{nm}-\upsilon^{y}_{mn}\upsilon^{x}_{nm})$ which is always semi-positive \cite{Sekh22}, the refined TDI then reduces to the following form:
\begin{align}
	\sum_{m\in occ}\sum_{n\in unocc}\frac{|\hbar\upsilon^{x}_{mn}|^{2}+|\hbar\upsilon^{y}_{mn}|^{2}- |-2\Im(\hbar\upsilon^{x}_{mn}\hbar\upsilon^{y}_{nm})|}{E^{2}_{mn}}\ge0.
\end{align}
Note that each term within the summation is positive. Now we multiply both sides by a factor, $c\delta(\hbar\omega+E_{mn})$, where $c=\pi e^2/\hbar$. This gives us an optical-like trace inequality:
\begin{align}
	\sum_{m\in occ}\sum_{n\in unocc}\frac{\Re\sigma^{xx}_{mn}(\mathbf{k},\omega)+\Re\sigma^{yy}_{mn}(\mathbf{k},\omega)}{|E_{mn}|}-\Im\overline{\sigma}^{xy}_{mn}(\mathbf{k},\omega)\ge0,
\end{align}
where we define $\Re\sigma^{aa}_{mn}(\mathbf{k},\omega)=c|\hbar\upsilon^{aa}_{mn}|^{2}\delta(\hbar\omega+E_{mn})/|E_{mn}|$, and $\Im\overline{\sigma}^{xy}_{mn}(\mathbf{k},\omega)=c|-2\Im(\hbar\upsilon^{x}_{mn}\hbar\upsilon^{y}_{nm}/E^{2}_{mn})|\delta(\hbar\omega+E_{mn})$. Note that the corresponding real or imaginary part of $\sigma^{ab}_{mn}(\mathbf{k},\omega)$ which are not considered in this study, can also be defined. Here, we introduce the notation for the real part because it is convenient for comparison and allows us to reduce the expression to the original form of optical conductivity easily. From the above equations, we obtain the upper limit on the MBC component at any interband transition contour in the first Brillouin zone:
\begin{align}
	\frac{\Re\sigma^{xx}_{mn}(\mathbf{k},\omega)+\Re\sigma^{yy}_{mn}(\mathbf{k},\omega)}{\hbar\omega}\ge\Im\overline{\sigma}^{xy}_{mn}(\mathbf{k},\omega).
\end{align}
Furthermore, by integrating over the momentum variable, we finally obtain the optical bound on the MBC at each frequency,
\begin{align}
	2\sum_{m\in occ}\sum_{n\in unocc}\int\frac{d^2\bf{k}}{(2\pi)^2}\frac{\Re\sigma^{xx}_{mn}(\mathbf{k},\omega)+\Re\sigma^{yy}_{mn}(\mathbf{k},\omega)}{\hbar\omega}\ge2\sum_{m\in occ}\sum_{n\in unocc}\int\frac{d^2\bf{k}}{(2\pi)^2}\Im\overline{\sigma}^{xy}_{mn}(\mathbf{k},\omega).
\end{align}
Here, we multiply both side of $2$ to maintain consistency with the definitions of quantum geometry and topology. Using simpler notation, we can rewrite it as
\begin{align}
	\frac{2[\Re\sigma_{xx}(\omega)+\Re\sigma_{yy}(\omega)]}{\hbar\omega}\ge2\Im\overline{\sigma}_{xy}(\omega).
\end{align}
where
\begin{align}
	\Re\sigma_{aa}(\omega)=\sum_{m\in occ}\sum_{n\in unocc}\int\frac{d^2\bf{k}}{(2\pi)^2}\Re\sigma^{aa}_{mn}(\mathbf{k},\omega), \quad\Im\overline{\sigma}_{xy}(\omega)=\sum_{m\in occ}\sum_{n\in unocc}\int\frac{d^2\bf{k}}{(2\pi)^2}\Im\overline{\sigma}^{xy}_{mn}(\mathbf{k},\omega).
\end{align}
Importantly, the left-hand term can be obtained through optical response measurements, whereas the right-hand term represents the corresponding imaginary part of the generalized optical Hall conductivity \cite{Ebert96,Oppeneer98,Gradhand13,Onishi24}.

\subsection{4: Souza-Wilkens-Martin sum rule and the relation between optical bound and quantum weight}
The SWM sum rule \cite{Souza00} can be derived directly by integrating the frequency of the real part of the optical conductivity. Let's start with the Kubo-Greenwood formula for optical conductivity: 
\begin{align}
	\sigma_{ab}(\omega)=\frac{e^{2}}{i\hbar}\sum_{m\ne n}\int\frac{d^2\bf{k}}{(2\pi)^2}\frac{f_{mn}(\mathbf{k})}{E_{mn}(\mathbf{k})}  
	\frac{\mathcal{M}^{a}_{mn}(\mathbf{k})\mathcal{M}^{b}_{nm}(\mathbf{k})}{\hbar\omega+E_{mn}(\mathbf{k})+i0^{+}},
\end{align}
where the velocity matrix element is given by $\mathcal{M}^{a}_{mn}(\mathbf{k})=\langle u_{m}(\mathbf{k})|\hbar\hat{\upsilon}_{a}|u_{n}(\mathbf{k})\rangle$, and the velocity operator is $\hat{\upsilon}_{a}=\frac{1}{\hbar}\frac{\partial \hat H}{\partial k_{a}}$. The Fermi-Dirac distribution function is given by $f_{n}(\mathbf{k})=1/[1+e^{(E_{n}(\mathbf{k})-\mu)/k_{B}T}]$. We denote $f_{mn}(\mathbf{k})=f_{m}(\mathbf{k})-f_{n}(\mathbf{k})$ and $E_{mn}(\mathbf{k})=E_{m}(\mathbf{k})-E_{n}(\mathbf{k})$, where $H(\mathbf{k})|u_{n}(\mathbf{k})\rangle=E_{n}(\mathbf{k})|u_{n}(\mathbf{k})\rangle$. The indices $a$ and $b$ represent the directional indices for current and incident electric field, respectively. Here, $\mu$ is the chemical potential. The interband contribution to the real part of the optical conductivity in the clean limit can be expressed in terms of the interband Berry connection (see the main context):
\begin{align}
	\Re\sigma_{aa}(\omega) 
	=\pi e^{2}\omega\sum_{m\ne n}\int\frac{d^2\bf{k}}{(2\pi)^2}
	r^{a}_{mn}r^{a}_{nm}\delta(\hbar\omega+E_{mn}(\bm{k})).
\end{align}
Here, we consider the case of zero temperature. Now, we multiply both sides of Eq. (S33) by a factor of $1/\omega$ and perform a frequency integration. Hence, we have
\begin{align}
	\int^{\infty}_{0}\frac{\Re\sigma_{aa}(\omega)}{\omega}d\omega&=\pi e^{2}\sum_{m\ne n}\int\frac{d^2\bf{k}}{(2\pi)^2}
	r^{a}_{mn}r^{a}_{nm}\int^{\infty}_{0}d\omega\delta(\hbar\omega+E_{mn}(\bm{k}))  \nonumber \\
	&=\frac{\pi e^{2}}{\hbar}\sum_{m\ne n}\int\frac{d^2\bf{k}}{(2\pi)^2}
	r^{a}_{mn}r^{a}_{nm}  \nonumber \\
	&=\frac{\pi e^{2}}{\hbar}\int\frac{d^2\bf{k}}{(2\pi)^2}
	g_{aa}(\mathbf{k}),
\end{align}
which is the SWM sum rule \cite{Souza00}, relating the optical weight to the quantum metric. In the second line, we use the delta function integration formula: $\int^{\infty}_{-\infty}d\omega f(\omega)\delta(\omega-a)=f(a)$. From the above delta function integration, we can see that all momenta in the Brillouin zone contribute to the integration. Note that negative frequencies have no contribution to the delta function integration.

According to the refined TDI, the integral of the quantum metric relates quantum geometry to topology. To reveal this relationship, we define a quantity called quantum weight \cite{Onishi24}, i.e.,
\begin{align}
	\mathcal{K}\equiv2\pi\int\frac{d^2\bf{k}}{(2\pi)^2}\mathrm{tr} g({\bf k}).
\end{align}
When considering the SWM sum rule for the two directions, we finally obtain
\begin{align}
	\frac{1}{2\pi}\int^{\infty}_{0}\frac{2[\Re\sigma_{xx}(\omega)+\Re\sigma_{yy}(\omega)]}{\omega}d\omega=\frac{e^{2}}{h}\mathcal{K}.
\end{align}
In the practical computation, we replace the $\omega$ in the denominator with $\hbar\omega$, i.e.,
\begin{align}
	\frac{1}{2\pi}\int^{\infty}_{0}\frac{2\hbar[\Re\sigma_{xx}(\omega)+\Re\sigma_{yy}(\omega)]}{\hbar\omega}d\omega=\frac{e^{2}}{h}\mathcal{K}.
\end{align}
We calculate $\Re\sigma_{aa}(\omega)$ in units of $e^2/h$, while the unit of $\hbar\omega$ is eV. In other words, the integral of the optical bound is equal to the quantum weight in units of $e^2/h$.

\subsection{5: Upper bound of the quantum weight and the optical gap}
Following the steps in Refs. \cite{Souza00,Onishi24}, we will obtain a new upper bound for the optical gap. Due to the semi-positive property of longitudinal optical conductivity, we always have the following inequality:
\begin{align}
	\frac{\hbar}{E_g}\int d\omega \Re\sigma_{aa}(\omega)\ge\int^{\infty}_{0}d\omega\frac{\Re\sigma_{aa}(\omega)}{\omega},
\end{align}
where $E_g$ is the optical gap \cite{Souza00,Onishi24}. On the other hand, there is the multi-band longitudinal f-sum rule \cite{Hazra19,Verma21,Mao23,Valderrama23}:
\begin{align}
	\int^{\infty}_{0}d\omega\Re\sigma_{aa}(\omega)=\frac{\pi e^2}{2}\sum_{m,n}\int \frac{d\mathbf{k}}{(2\pi)^2}[M^{-1}_{mn}(\mathbf{k})]_{aa}\langle c^{\dagger}_{\mathbf{k}m}c_{\mathbf{k}n}\rangle,
\end{align}
where the inverse mass tensor is given by $[M^{-1}_{mn}(\mathbf{k})]_{aa}=\sum_{\alpha\beta}U^{\dagger}_{m,\alpha}(\mathbf{k})\frac{\partial^2H_{\alpha\beta}(\mathbf{k})}{\partial(\hbar k_a)^2}U_{\beta,n}(\mathbf{k})$.
Here, $\alpha$ and $\beta$ label the internal degrees of freedom, such as orbital, site, and spin. $H_{\alpha\beta}(\mathbf{k})$ is the matrix element of the non-interacting Hamiltonian. We only consider the non-interacting case, which leads to a greatly simplified form. Combining this with the integrated TDI,
\begin{align}
	\mathcal{K}\ge2\mathcal{V}\ge\sum^{n_t}_{s=1}|\mathcal{N}_{s}|,
\end{align}
we have derived a longer inequality:
\begin{align}
	\frac{\pi\hbar^2}{E_{g}}\sum_{n}\int \frac{d\mathbf{k}}{(2\pi)^2}([M^{-1}_{nn}(\mathbf{k})]_{xx}+[M^{-1}_{nn}(\mathbf{k})]_{yy})f_n({\mathbf{k}})\ge\mathcal{K}\ge2\mathcal{V}\ge\sum^{n_t}_{s=1}|\mathcal{N}_{s}|,
\end{align}
which shows that a larger $E_g$ implies a tighter difference between each quantity if the inverse mass tensor has only slight variations. We can also obtain an upper bound for the optical gap:
\begin{align}
	\frac{\pi\hbar^2}{2\mathcal{V}}\sum_{n}\int \frac{d\mathbf{k}}{(2\pi)^2}([M^{-1}_{nn}(\mathbf{k})]_{xx}+[M^{-1}_{nn}(\mathbf{k})]_{yy})f_n({\mathbf{k}})\ge E_g.
\end{align}
If the non-interacting Hamiltonian is given by $\hbar^2\mathbf{k}^2/(2m_e)$ \cite{Kubo57,Onishi24}, it can be simplified to:
\begin{align}
	\frac{\pi\hbar^2n_e}{2m_e\mathcal{V}}\ge E_g,
\end{align}
where $n_e$ is the electron density. In other words, particle density, electron mass, and quantum volume directly constrain the maximum size of the optical gap.

\subsection{6: Numerical results on band structure and quantum weight during the topological phase transition for the three models}
In this section, we present numerical results to better understand the physical properties of topological phase transitions (see Fig. S1 and S2), as well as the crucial inequality $2\sqrt{\det g(\mathbf{k})}\ge\overline{\overline{\Omega}}_{ab}(\mathbf{k})\ge\sum^{\text{occ}}_{n=1}|\Omega^{n}_{ab}(\mathbf{k})|$ (see Fig. S3), for the three models discussed in the main text. All figures use the same parameters as those in Figs. 1 and 2 (see the main text), except for the Kane–Mele model, where $(t,\lambda_{SO},\lambda_{R})=(1,0.06t,0.05t)$, and for the HOTI model, where we do not include the mirror-symmetry-breaking term.

\begin{figure}[]
	\includegraphics[width=1.0\textwidth]{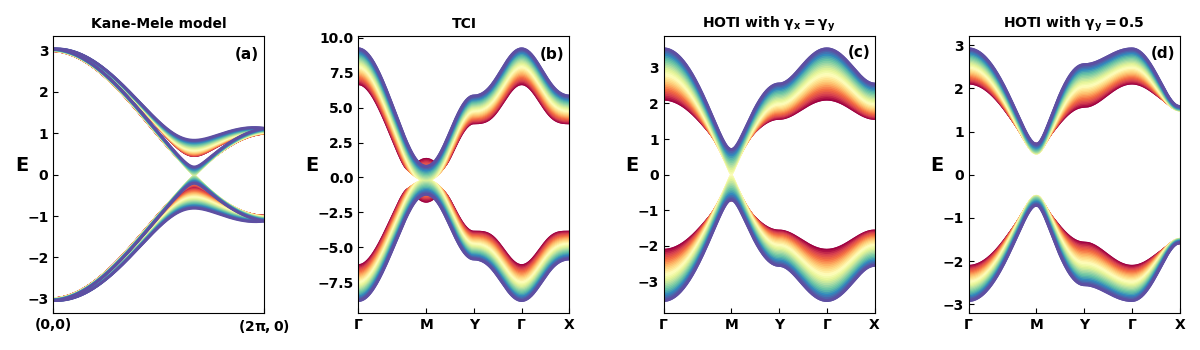}
	\caption{Band structure of the three models as a function of $m$ along the selected Brillouin zone path near the topological phase transition. (a) $m\in(0,0.5)$. (b) $m\in(2.5,5)$. (c,d) $m\in(0.5,1.5)$.}
\end{figure}

\begin{figure}[]
	\includegraphics[width=1.0\textwidth]{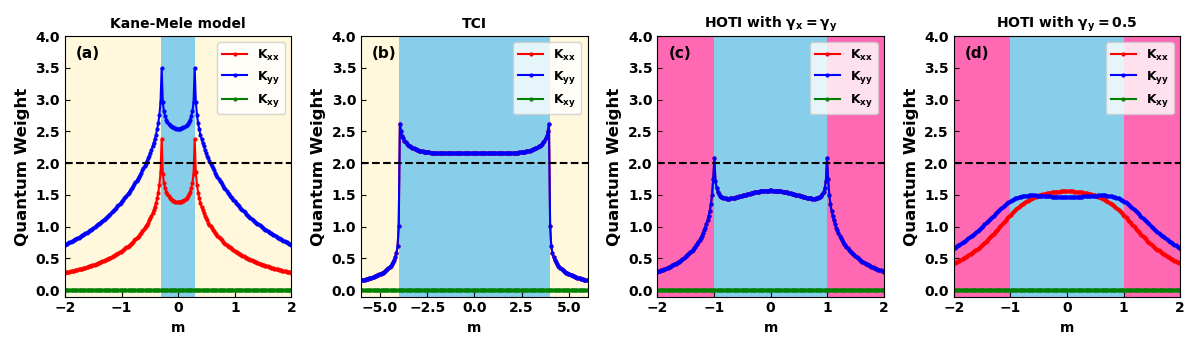}
	\caption{Optical weight of the three models as a function of 
		$m$. The figures clearly show two types of behavior of the quantum weight: $K_{xx}\ne K_{yy}$ (a, d) and $K_{xx}=K_{yy}$ (b, c) along the $x$ and $y$ axes. In panels (a, b, c), the optical weight exhibits discontinuous behavior at the critical point, while in panel (d), the transition is smooth, corresponding to another type of topological phase transition without gap closing. Here, the optical weight is given by $K_{ab}\equiv2\pi\int\frac{d^2k}{(2\pi)^2}g_{ab}({\bf k})$.}
\end{figure}

\begin{figure}[]
	\includegraphics[width=0.99\textwidth]{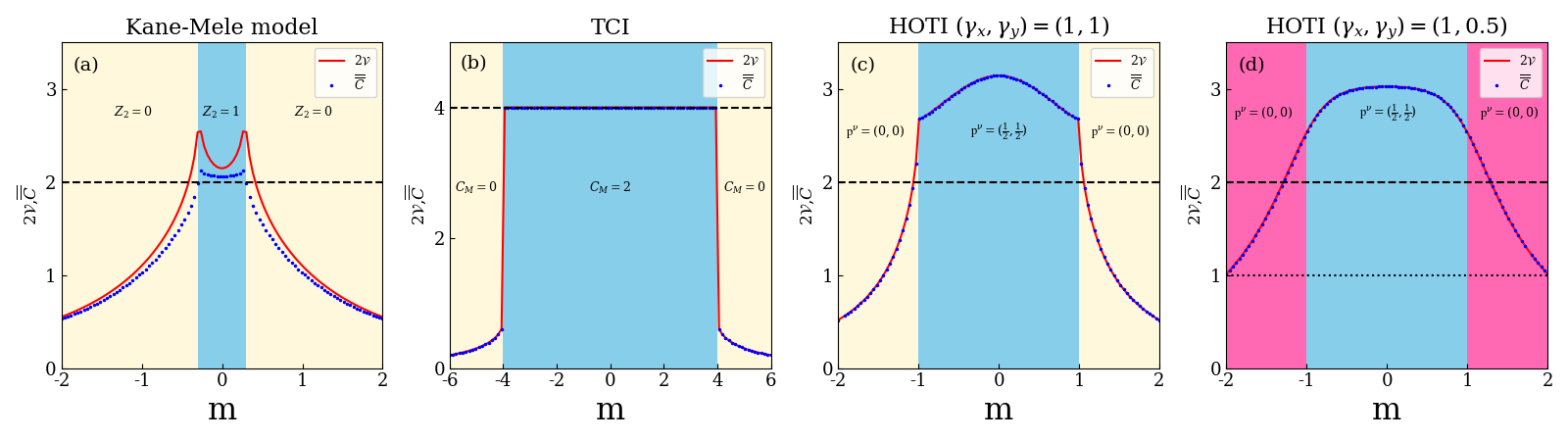}
	\caption{The double quantum volume $(2\mathcal{V})$ and "pseudo-Chern number" of the projection form of MBC $(\overline{\overline{C}})$ as a function of $m$ for the Kane-Mele model (a), TCI (b), and HOTI (c,d). The black dashed and dotted lines represent the number of boundary states. The remaining parameters are the same as in Fig. 1 and Fig. 2. We can see that the double quantum volume and the "pseudo-Chern number" of the projection form of MBC are equal in the case of TCI and HOTI. Note that we add a small term $\delta \text{I}$ to $iP(\mathbf{k})[\partial_a P(\mathbf{k}),\partial_b P(\mathbf{k})]$ to stabilize the computation of $\overline{\overline{C}}$, where $\text{I}$ is an identity matrix and $\delta$ is a small number, say $0.00001$.}
\end{figure}


\end{document}